\documentclass[aps,amsmath,twocolumn,amssymb,floatfix,showpacs,superscriptaddress,nofootinbib,longbibliography]{revtex4-1}
\usepackage{MnSymbol}
\usepackage{braket}
\usepackage[dvipsnames]{xcolor}
\usepackage{float}
\usepackage{subfigure}
\usepackage{tikz}
\usepackage[colorlinks=true,linktoc=page,linkcolor=magenta,citecolor=blue,urlcolor=purple]{hyperref}

\newcommand{\vect}[1]{\boldsymbol{\mathrm{#1}}}
\newcommand{\Eq}[1]{Eq.~(\ref{#1})}
\mathchardef\mhyphen="2D 

\newcommand{\etal}{{\it et al.~}}
\newcommand{\ie}{{i.e.,\,\,}}
\newcommand{\eg}{{e.g.,~}}
\newcommand{\ua}{{\uparrow }}
\newcommand{\da}{{\downarrow }}

\newcommand\bea{\begin{eqnarray}}
	\newcommand\eea{\end{eqnarray}}
\newcommand\beq{\begin{equation}}  
	\newcommand\eeq{\end{equation}}

\newcommand{\non}{\nonumber}  
\usepackage[normalem]{ulem}
\definecolor{lime}{HTML}{A6CE39}
\usepackage{sidecap,tikz}
\DeclareRobustCommand{\orcidicon}{\hspace{-1.0mm}
	\begin{tikzpicture}
		\draw[lime, fill=lime] (0.0,0.0) 
		circle [radius=0.15] 
		node[white] {{\fontfamily{qag}\selectfont \tiny \,ID}};
		\draw[white, fill=white] (-0.0525,0.095) 
		circle [radius=0.007];
	\end{tikzpicture}
	\hspace{-3.0mm}
}
\foreach \x in {A, ..., Z}{\expandafter\xdef\csname orcid\x\endcsname{\noexpand\href{https://orcid.org/\csname orcidauthor\x\endcsname}{\noexpand\orcidicon}}
}

\AtBeginDocument{%
	\newwrite\bibnotes
	\def\bibnotesext{Notes.bib}
	\immediate\openout\bibnotes=\jobname\bibnotesext
	\immediate\write\bibnotes{@CONTROL{REVTEX41Control}}
	\immediate\write\bibnotes{@CONTROL{%
			apsrev41Control,author="08",editor="1",pages="1",title="1",year="1"}}
	\if@filesw
	\immediate\write\@auxout{\string\citation{apsrev41Control}}%
	\fi
}%

\allowdisplaybreaks

\begin{document}

\title{Generation of higher-order topological insulators using periodic driving}  

\author{Arnob Kumar Ghosh\orcidA{}}
\email{arnob.ghosh@physics.uu.se}
\affiliation{Institute of Physics, Sachivalaya Marg, Bhubaneswar-751005, India}
\affiliation{Homi Bhabha National Institute, Training School Complex, Anushakti Nagar, Mumbai 400094, India}
\affiliation{Department of Physics and Astronomy, Uppsala University, Box 516, 75120 Uppsala, Sweden}

\author{Tanay Nag\orcidB{}}
\email{tanay.nag@hyderabad.bits-pilani.ac.in}
\affiliation{Department of Physics and Astronomy, Uppsala University, Box 516, 75120 Uppsala, Sweden}
\affiliation{Department of Physics, BITS Pilani-Hyderabad Campus, Telangana 500078, India}

\author{Arijit Saha\orcidC{}}
\email{arijit@iopb.res.in}
\affiliation{Institute of Physics, Sachivalaya Marg, Bhubaneswar-751005, India}
\affiliation{Homi Bhabha National Institute, Training School Complex, Anushakti Nagar, Mumbai 400094, India}

\begin{abstract}
Topological insulators~(TIs) are a new class of materials that resemble ordinary band insulators in terms of a bulk band gap but exhibit protected metallic states on their boundaries. In this modern direction, higher-order TIs~(HOTIs) are a new class of TIs in dimensions $d>1$. These HOTIs possess $(d - 1)$-dimensional boundaries that, unlike those of conventional TIs, do not conduct via gapless states but are themselves TIs. Precisely, an $n^{\rm th}$ order $d$-dimensional higher-order topological insulator is characterized  by the presence of boundary modes that reside on its $d_c=(d-n)$-dimensional boundary. For instance, a three-dimensional second (third) order TI hosts gapless (localized) modes on the hinges (corners), characterized by $d_c = 1 (0)$. Similarly, a second-order TI in two dimensions only has localized corner states ($d_c = 0$). These higher-order phases are protected by various crystalline as well as discrete symmetries.
The non-equilibrium tunability of the topological phase has been a major academic challenge where periodic Floquet drive provides us golden opportunity to overcome that barrier.
Here, we discuss different periodic driving protocols to generate Floquet higher-order TIs while starting from a non-topological or first-order topological phase. Furthermore, we emphasize that one can generate the dynamical anomalous $\pi$-modes along with the concomitant $0$-modes. The former can be realized only in a dynamical setup. We exemplify the Floquet higher-order topological modes 
in two and three dimensions in a systematic way.  Especially, in two dimensions, we demonstrate a Floquet second-order TI hosting $0$- and $\pi$ corner modes. Whereas a three-dimensional Floquet second-order TI and Floquet third-order TI manifest one- and zero-dimensional hinge and corner modes, respectively.
\end{abstract}

\maketitle

\section{Introduction}

The advent of the integer quantum Hall effect~(IQHE) by Klaus von Klitzing in 1980~ \cite{KlitzingPRL1980} introduces the notion of topology in the field of condensed matter physics. 
Soon after the discovery of IQHE, the topological characterization of the quantum Hall states (QHS) 
is predicted by Thouless, Kohmoto, Nightingale, and den Nijs, in terms of the quantized Hall conductance [TKNN invariant] also known as Chern number~\cite{TKNNPRL1982}. The QHS hosts gapless chiral edge modes only at the boundaries of the system. The topological non-triviality in the QHS originates from the fact that the appearance of the boundary modes does not depend on the minute details of the system; instead, one continues to observe these modes as long as an energy gap sustains between the consecutive Landau levels.
Classically, one may understand the origin of the edge-states from the skipping motions of the electrons along the edges of the syste~\cite{hasan2010colloquium}.

The generation of QHS depends upon the externally-applied high magnetic field, which breaks discrete time-reversal symmetry~(TRS). In 1988, Duncan Haldane, in his seminal paper, proposed an elegant alternative way to realize IQHE employing two-dimensional~(2D) hexagonal lattice without any net magnetic flux per unit cell~\cite{HaldanePRL1988}. The TRS breaking mechanism in this hexagonal setup is engineered by introducing an imaginary second nearest-neighbor hopping term. Employing Bloch band theory, one can find out the band structure of the system, and it turns out that the system 
is insulating in bulk. However, when a boundary is imposed on the system \ie considering the 
finite size of the system along at least one direction, it exhibits gapless chiral edge states. Also, the topological nature of the non-trivial bulk bands can be characterized by a non-zero TKNN invariant/ Chern number~\cite{HaldanePRL1988}. Due to the chiral nature, the edge modes can propagate only along one direction. Thus, these edge-states are robust against disorder as back-scattering is prohibited due to the unavailability of oppositely moving states. This phenomenon of generating non-trivial TRS-breaking topological states without an external magnetic field is also known as the quantum anomalous Hall~(QAH) effect~\cite{ChangQAHRMP2023}.

During the past two decades, researchers in this area started asking the interesting question of 
realizing a quantum Hall-like state without breaking TRS. In this direction, Kane and Mele~\cite{kane2005quantum,kane2005quantumPRL,BernevigBOOK,GruznevNanoLetter2018}, and Bernevig, Hughes and Zhang~\cite{BHZPRL2006,BernevigBOOK} independently proposed the elegant idea of quantum spin Hall~(QSH) effect. In a QSH insulator~(QSHI) or 2D topological insulator~(TI), the spin-orbit coupling~(SOC) plays the role of the magnetic field that is momentum dependent. However, SOC 
does not break TRS. Moreover, the QSHI exhibits spin currents instead of charge currents in terms of quantized spin-Hall conductance. Due to the presence of TRS, it is obvious to have Kramers' degeneracy. Thus, a QSHI exhibits two counter-propagating edge modes per edge \ie two opposite spins propagate along opposite directions (left mover and right mover), and this phenomenon is known as spin-momentum locking~\cite{qi2011topological}. As a result, the total charge current in a 2D TI becomes zero. However, one can still obtain a quantized spin current, which is the QSH effect~\cite{ZhangSpincurrent2006}.

Soon after the discovery of 2D TI, the bulk-boundary correspondence~(BBC) is generalized for a three-dimensional~(3D) system, and the idea of a 3D TI is formulated~\cite{fu2007topological,MoorePRB2007,RahulRoyPRB2009,hasan2010colloquium}. In three dimensions, the TI exhibits a 2D surface state with gapless Dirac cones while the bulk remains insulating. Theoretically, Bi$_{1-x}$Sb$_x$, $\alpha$-Sn and HgTe under uniaxial strain, Bi$_2$X$_3$ (X=Se,Te), Sb$_2$Te$_3$, etc. are proposed to be the probable material platform to manifest 3D TI~\cite{hasan2010colloquium,FuKane2007,Zhang2009}. A few experimental observations have been put forward illustrating the evidence of 3D TI hosting gapless surface Dirac cones in  Bi$_{1-x}$Sb$_x$~\cite{hsieh2008topological}, Bi$_2$X$_3$ (X=Se,Te)~\cite{Xiaexperiment2009,chen2009experimental}, etc. In three dimensions, however, there is a possibility of realizing two kinds of TIs- strong TI with an odd number of Dirac cones on the surface and weak TI with an even number of Dirac cones on the surface~\cite{fu2007topological}. The weak TIs are not robust against disorder due to the possibility of inter-node scattering and as such are equivalent to the band insulators~\cite{fu2007topological}.

The breaking of TRS in QHS facilitates the computation of the TKNN invariant or the Chern number 
($C$) for topological characterization of such system~\cite{TKNNPRL1982}. However, when the TRS is preserved, the total Chern number vanishes~\cite{FuTimereversal2006}. Thus, the Chern number can not be employed to characterize the QSH phase, which preserves TRS. Although one may still be able to define the spin Chern number $C_{\rm S}$ if the $z$-component of the spin (${\rm S}_z$) still remains preserved such that $C_{\rm S}=(C_\uparrow-C_\downarrow)/2$; with $C_\uparrow$~($C_\downarrow$) representing the Chern number of the up (down) spin-sector. However, when ${\rm S}_z$ is not preserved, 
one can, however, find out a $\mathbb{Z}_2$ topological chatacterization~\cite{FuTimereversal2006,kane2005quantumPRL,FuKane2007,fu2007topological,MoorePRB2007,BernevigBOOK,RahulRoyPRB2009b}. The $\mathbb{Z}_2$-invariant is computed employing the time-reversal polarization of the bulk bands~\cite{FuTimereversal2006}. For 2D TI, the $\mathbb{Z}_2$ index $\nu_0$ takes the values $0$ and $1$ for the topological and the non-topological case, respectively~\cite{FuTimereversal2006,kane2005quantumPRL,FuKane2007,BernevigBOOK}. In three dimensions, however, one needs four $\mathbb{Z}_2$ indices: $(\nu_0;\nu_1,\nu_2,\nu_3)$ to fully characterize the system~\cite{fu2007topological,MoorePRB2007}. Here, $\nu_0=1$ implies a strong topological phase, and the surface states accommodate odd number of Dirac cones. However, 
$\nu_0=0$ can indicate a trivial or weak-topological phase hosting even number of Dirac cones. In particular, strong topological phase (weak topological phase) is characterized by $\nu_0=1~(\nu_0=0)$, 
once at least one of the $\nu_{1,2,3}$'s remains non-zero. Trivial phase is designated by $\nu_{1,2,3}=0$, along with $\nu_0=0$.  The latter exhibits no/ gapped surface states.
Overall, the so far discussed 2D and 3D TIs are referred to as first-order TIs~(FOTIs).

The FOTIs are protected by TRS, and one can adiabatically connect the TIs to atomic insulators only if TRS is explicitly broken or the bulk gap is closed. Thus, only the TRS plays a pivotal role in BBC in the case of a FOTI. However, the advent of topological crystalline insulator~(TCI) establishes the role of 
spatial symmetries (\eg space-inversion, mirror, etc.) in the BBC~\cite{fu2011topological,TurnerCrystallinePRB2012,RyuPRB2013,MorimotoPRB2013,SatoCrystallinePRB2014,AndoARCMP2015,Hsu_2019}.
Here, SOC is not that much necessary to procure the topological phase. The BBC in a TCI is more indeterminate and depends on the information about the boundary termination. Moreover, the boundary may possess lower symmetries compared to the bulk in a TCI, and the surfaces/edges that satisfy the crystalline symmetry requirements host gapless states. Nevertheless, TCIs are robust against a symmetry-preserving disorder that does not close the bulk gap. From the experimental point of view, mirror-symmetry protected TCI has been detected in materials like Pb$_{1-x}$Sn$_x$Te~\cite{Xucrystalline2012}, Pb$_{1-x}$Sn$_x$Se~\cite{Dziawacrystalline2012}, etc.

\begin{figure}[]
    \centering
    \subfigure{\includegraphics[width=0.47\textwidth]{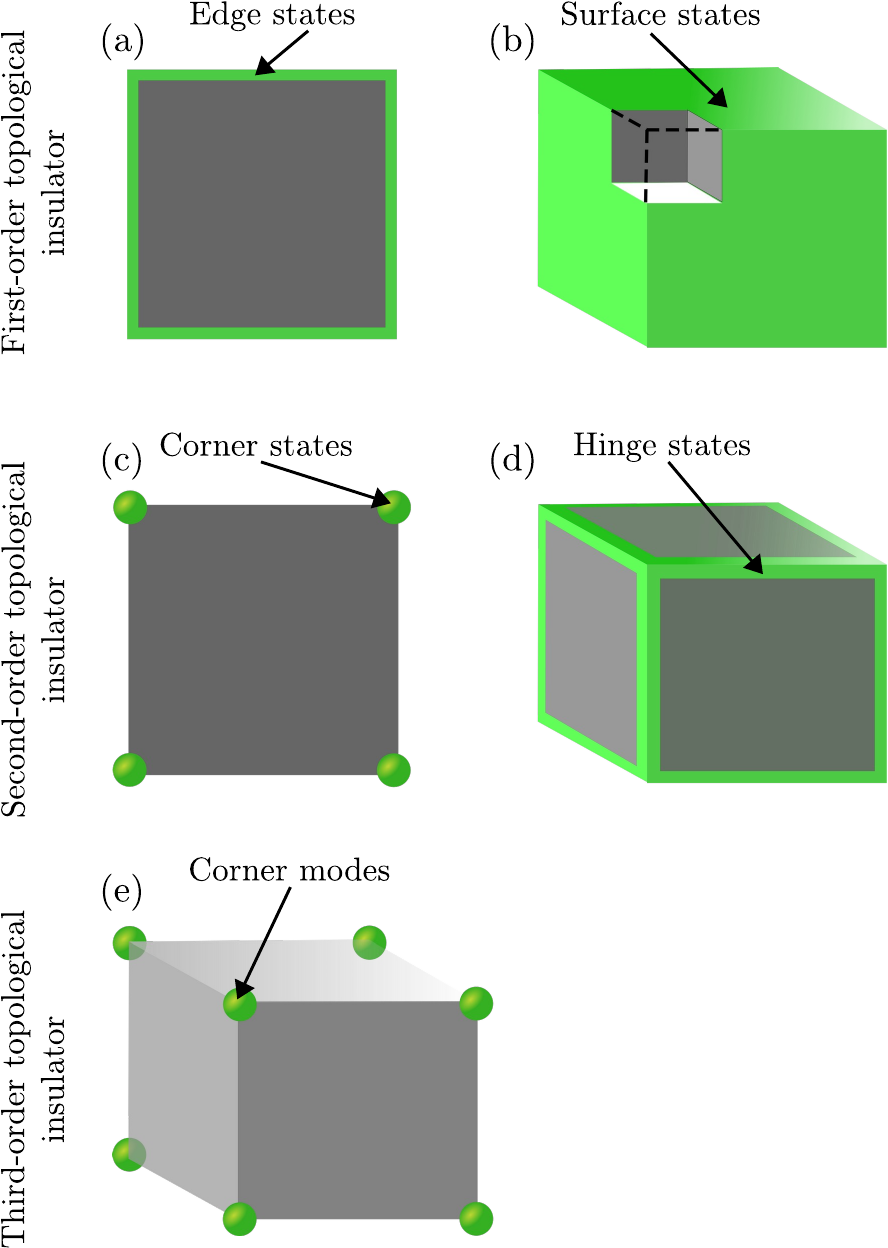}}
    \caption{Schematics of $n^{\rm th}$ order TIs are illustrated. Panels (a) and (b) represent a 2D and 3D TI, respectively. Both systems have gapped-bulk states [gray region]. However, a 2D~(3D) TI exhibit 1D~(2D) edge states [represented by green lines]~(surface states [represented by green surfaces]). Schematics of 2D and 3D SOTI are depicted in panels (c) and (d). Both systems have a gapped bulk and $(d-1)$-dimensional boundary states [gray regions]. However, a 2D~(3D) SOTI exhibit 0D~(1D) corner states [represented by green spheres]~(hinge states [represented by green lines]). (e) Schematic of a 3D TOTI is demonstrated. The system hosts a gapped bulk as well as $(d-1)$- and $(d-2)$-dimensional boundary states [gray regions]. However, a TOTI is accompanied by 0D localized corner states [represented by green spheres].
	}
	\label{static}
\end{figure}

Very recently, the concept of BBC has been transcended to a new class of topological materials called the higher-order TI~(HOTI)~\cite{benalcazar2017,benalcazarprb2017,Song2017,Langbehn2017,schindler2018,Franca2018,Ezawaphosphorene,wang2018higher,Ezawakagome,Geier2018,Khalaf2018,ezawa2019second,luo2019higher,Roy2019,RoyGHOTI2019,Trifunovic2019,agarwala2019higher,Dutt2020,Szumniak2020,Ni2020,Costanpj2021,BiyeXie2021,trifunovic2021higher,YangCommun2023}, where the spatial and non-spatial (time-reversal, particle-hole, chiral) symmetries come together to protect such phase. In particular, a $d$-dimensional HOTI of order $n$, like the FOTI, possesses a gapped bulk states [see Fig.~\ref{static}], however unlike FOTI, they do not manifest $(d-1)$-dimensional gapless boundary states, rather a $(d-n)$-dimensional boundary modes [see Figs.~\ref{static} (c)-(e)]. Precisely, a 2D second-order TI~(SOTI) exhibits localized $0$-dimensional~(0D) corner states and gapped edge states [see Fig.~\ref{static} (c)]. At the same time, a 3D SOTI is characterized by the presence of gapless, dispersive 1D hinge states and gapped surface states [see Fig.~\ref{static} (d)]. In contrast, a 3D third-order TI~(TOTI) displays localized 0D corner states while 
the surfaces and the hinges remain gapped [see Fig.~\ref{static} (e)]. Thus, the materials previously thought to be topologically trivial due to the absence of $(d-1)$-dimensional boundary states may turn out to be HOTI, thereby enhancing the quest for searching new topological materials. The higher-order polarization, like quadrupole and octupole moments, can be formulated for the topological characterization of HOTIs~\cite{benalcazar2017,benalcazarprb2017}. In this intriguing direction, a few experimental proposals have also been put forward employing solid-state systems~\cite{schindler2018higher,Experiment3DHOTI.VanDerWaals,Aggarwal2021,ShumiyaHOTI2022,LeeNatureComm2023}, phononic crystals~\cite{serra2018observation}, acoustic systems~\cite{xue2019acoustic,ni2019observation,Experiment3DHOTI.aSonicCrystals,Ni2020}, electric-circuit setups~\cite{imhof2018topolectrical}, photonic lattice~\cite{PhotonicChen,PhotonicXie,mittal2019photonic} etc. Thus, such higher-order systems constitute a distinctive new family of topological phases of matter.


In recent times, light-matter interaction has become a fascinating research direction from both theoretical and experimental perspectives. The application of light in a solid-state system can architect light-induced insulator-metal transition~\cite{Metalinsulatorlight1998}, light-induced photovoltaic effect~\cite{oka09photovoltaic,Leephotovoltic2008}, photo-thermoelectric effect~\cite{XiaodongPhoto-Thermoelectric2010}, light-induced superconductivity~\cite{FaustiScience}, etc. Moreover, the non-equilibrium generation of topological states of matter has become another intriguing research direction since the last decade~\cite{Basov2017,oka09photovoltaic,Kitagawacharacterization2010,lindner11floquet,FloquetGuPRL2011,Rudner2013,Piskunow2014,Usaj2014,Nathan_2015,Eckardt2015,MikamiBW2016,Yan2017,Eckardt2017,oka2019,HeNatCommun2019,NHLindner2020,Bao2022}. In a time-periodic system, Floquet theory provides the prescription for analyzing the non-equilibrium system with the concept of quasi-energy and quasi-states~\cite{FLoquetpaper1883}, and thus the periodically driven systems are also called the Floquet systems. In a static equilibrium system, there are only a few ways to tune the topological properties of a system, e.g., by changing the width of the system~\cite{bernevig2006quantum,konig2007quantum}, doping concentration, etc. However, Floquet engineering provides us with the on-demand control of the topological properties of a system in the presence of an external periodic drive~\cite{Basov2017,oka09photovoltaic,Kitagawacharacterization2010,lindner11floquet,kitagawa11transport,FloquetGuPRL2011,Dora2012,Rudner2013,Thakurathi2013,Piskunow2014,Usaj2014,BenitoPRB2014,Nathan_2015,Eckardt2015,sacramento2015,MikamiBW2016,Yan2017,Eckardt2017,oka2019,NHLindner2020,rxzhangPRL2021,Bao2022,MondalPRB2023,mondal2023engineering}. Employing Floquet engineering, one can generate the topological phase starting from a topologically trivial system. The resulting BBC here becomes intriguing in the presence of the extra-temporal dimension. Another intriguing aspect of Floquet engineering is that one can also engineer the closing and reopening of bulk gaps at the Floquet-zone boundary, i.e., at quasienergy $\omega/2$ when $\omega$ lies within the intermediate regime (\ie $\omega \sim$ bandwidth of the system); with $\omega$ being the driving frequency. Thus within the intermediate frequency regime, there is a possibility of realizing anomalous boundary modes at finite quasienergy, namely $\pi$-modes, with concurrent regular $0$-modes~\cite{Kitagawacharacterization2010,JiangColdAtomPRL2011,Rudner2013,Piskunow2014,Usaj2014,Yan2017,Eckardt2017,NHLindner2020}. These $\pi$-modes do not exhibit any static analog and, thus, true dynamical in nature. The prodigious experimental development of Floquet systems based on solid-state setup~\cite{WangScience2013,Mahmood2016,McIver2020}, ultra-cold atoms~\cite{Jotzu2014,Wintersperger2020}, acoustic systems~\cite{Peng2016,fleury2016floquet}, photonic platforms~\cite{RechtsmanExperiment2013,Maczewsky2017}, etc., add further merits to this field towards their realization and possible device application. However, the light-induced quantum phenomena are truly non-equilibrium in nature, and the physical signatures of dynamical modes, as well as the stabilization of these systems, are not very clear as of yet~\cite{kitagawa11transport,NHLindner2020}, and most of the understanding is derived from the transport measurements~\cite{kitagawa11transport}. 

The remainder of the review article is organized as follows. In Sec.~\ref{sec2}, we present the basic models of static HOTI and their basic phenomelology. In Sec.~\ref{sec3}, we present a primer on the Floquet theory and discuss the Floquet FOTI based on driven Bernevig-Hughes-Zhang~(BHZ) model. Sec.~\ref{Sec:FloquetHOTI} is devoted to the detailed discussion of Floquet generation HOTIs in two 
and three dimensions via different periodic driving. We present a discussion and possible outlook in Sec.~\ref{sec5}. The current experimental progress for the realization of the HOTI 
phase is discussed in Sec.~\ref{sec6}. Finally, we summarize and conclude our article in Sec.~\ref{sec7}.
\section{Static HOTI models} \label{sec2}
In this section, we discuss the features of a few static HOTI models before moving toward the non-equilibrium generation of HOTI.
\subsection{2D SOTI}
In two dimensions, the most popular models for describing the SOTI phase are the Benalcazar-Bernevig-Hughes (BBH) model~\cite{benalcazar2017,benalcazarprb2017} and BHZ model with a four-fold rotation $C_4$ and TRS $\mathcal{T}$ breaking Wilson-Dirac~(WD) mass term~\cite{schindler2018,Roy2019}. However, one can obtain a unitary transformation to unearth a one-to-one connection between these two models~\cite {trifunovic2021higher}. Here, we briefly discuss these two models' Hamiltonians, their symmetries, topological phase boundaries, and topological characterization.
\subsubsection{Model-1: The BBH model} \label{Subsection:BBHmodel}

\begin{figure}[]
	\centering
	\includegraphics[width=0.3\textwidth]{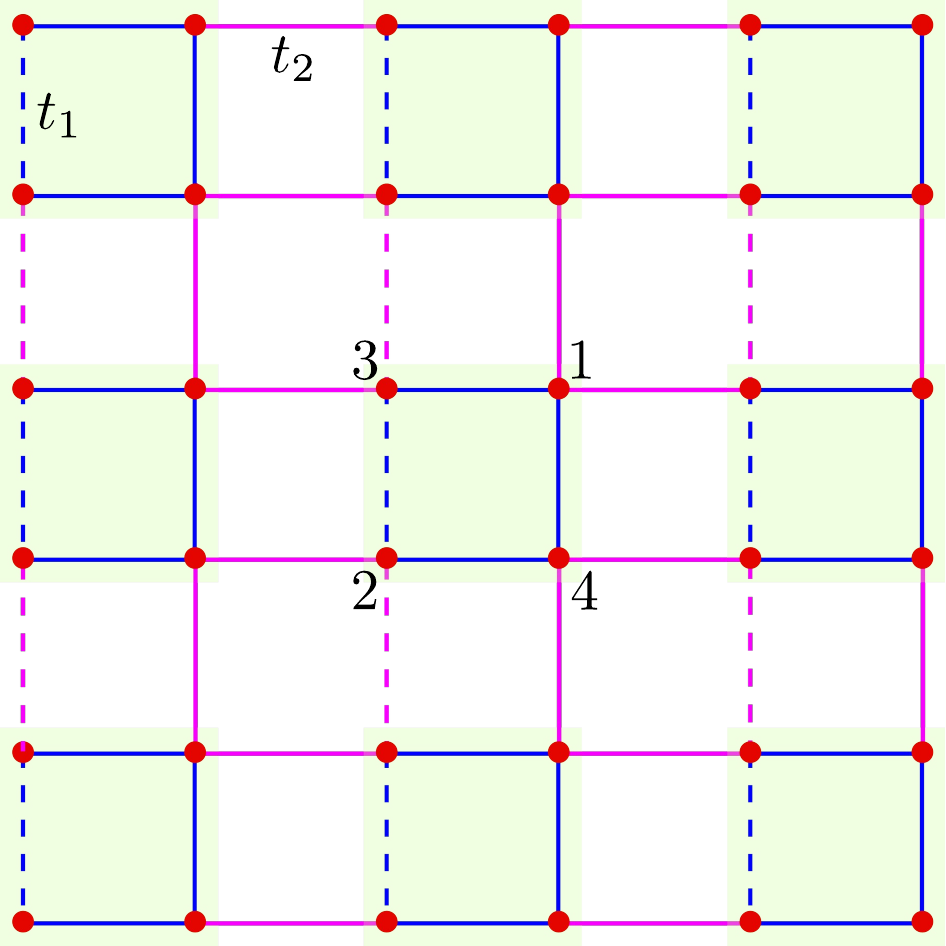}
	\caption{A schematic representation of the BBH model is demonstrated. The unit cell comprises of four pseudo-spin/orbital degrees of freedom, represented by 1-4. Here, blue and magenta solid lines represent intra- and inter-cell hoppings, respectively. At the same time, the dashed lines portray a negative hopping value.
	}
	\label{BBH_Sch}
\end{figure}

The BBH model is based on a four-band Bloch Hamiltonian, which reads as~\cite{benalcazar2017,benalcazarprb2017}
\begin{align}
	\mathcal{H}_{\rm BBH} (\vect{k}) =& \left[ t_1 + t_2 \cos k_x \right] \tau_x \sigma_0 - t_2 \sin k_x  \ \tau_y \sigma_z \non \\
	&- \left[ t_1 + t_2 \cos k_y \right] \tau_y \sigma_y - t_2 \sin k_y \ \tau_y \sigma_x \ ,
	\label{BBHModel}
\end{align}
where the Pauli matrices $\vect{\tau}$ and $\vect{\sigma}$ act on two different pseudo-spin/orbital degrees of freedom. Here, $t_1$ and $t_2$ represent the intra- and inter-cell hopping amplitudes, respectively. The lattice representation of the BBH model is schematically depicted in Fig.~\ref{BBH_Sch}. Below, we discuss the various symmetries that the bulk Hamiltonian $\mathcal{H}_{\rm BBH} (\vect{k})$ respects along with the corresponding symmetry operations:
\begin{itemize}
	\item TRS with $\mathcal{T}=\tau_0 \sigma_0 \mathcal{K}$: $\mathcal{T} \mathcal{H}(\vect{k}) \mathcal{T}^{-1}=\mathcal{H}(-\vect{k})$; with $\mathcal{K}$ being the complex-conjugation operator,
	\item Charge-conjugation symmetry with $\mathcal{C}=\tau_z \sigma_0 \mathcal{K}$:  $\mathcal{C} \mathcal{H}(\vect{k}) \mathcal{C}^{-1}= -\mathcal{H}(-\vect{k})$,
	\item Sublattice or chiral symmetry with $\mathcal{S}=\tau_z \sigma_0$:  $\mathcal{S} \mathcal{H}(\vect{k}) \mathcal{S}^{-1}= -\mathcal{H}(\vect{k}) $,
	\item Mirror symmetry along $x$ with $\mathcal{M}_x= \tau_x \sigma_z$: $\mathcal{M}_x \mathcal{H}(k_x,k_y) \mathcal{M}_x^{-1}= \mathcal{H}(-k_x,k_y) $,
	\item Mirror symmetry along $y$ with $\mathcal{M}_y= \tau_x \sigma_x$: $\mathcal{M}_y \mathcal{H}(k_x,k_y) \mathcal{M}_y^{-1}= \mathcal{H}(k_x,-k_y) $,
\end{itemize}
Here, we have removed the subscript of the Hamiltonian $\mathcal{H}_{\rm BBH} (\vect{k})$ for brevity.

\begin{figure}[]
	\centering
	\includegraphics[width=0.49\textwidth]{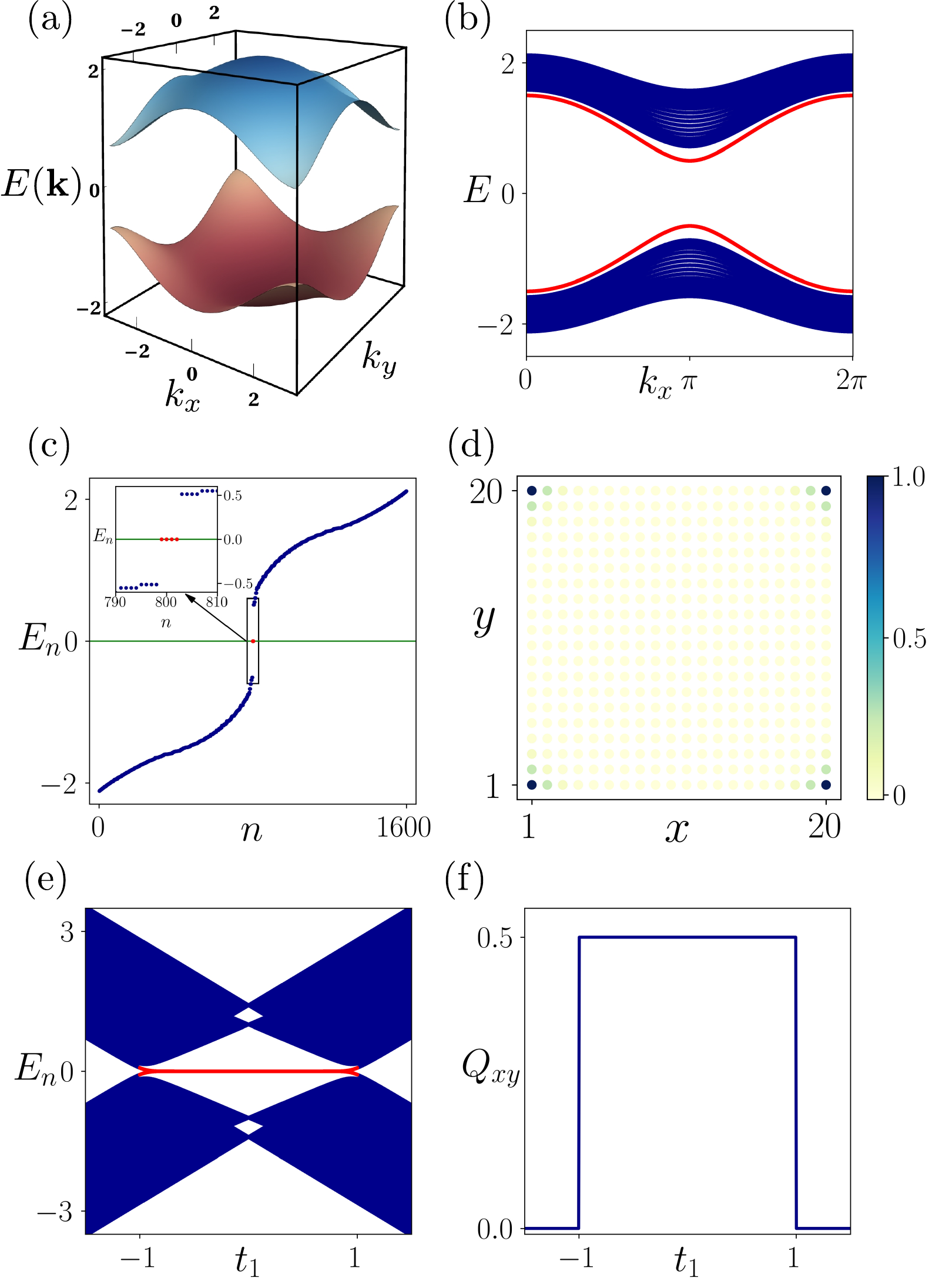}
	\caption{(a) We depict the bulk bands of the BBH Hamiltonian $\mathcal{H}_{\rm BBH} (\vect{k})$ [Eq.~(\ref{BBHModel})] as a function of the crystal momenta $k_x$ and $k_y$. In panel (b), we illustrate the slab geometry eigenvalue spectrum of the BBH model Hamiltonian, considering 20 lattice sites in the $y$-direction. Both the bulk and edge exhibit a finite gap, which is necessary for a SOTI phase. In panel (c), the eigenvalue spectrum of a finite size ($20 \times 20$ lattice sites) BBH model is demonstrated. The zero-energy eigenvalues are shown in the inset. The LDOS corresponding to the zero-energy eigenstates is depicted as a function of lattice geometry $x$ and $y$ in panel (d). It is evident that the zero-energy states are localized at the corners of the system. In panel (e), we illustrate the eigenvalue spectra of the finite-size Hamiltonian as a function of the intra-cell hopping amplitude $t_1$. The zero-energy corner states are obtained for $\lvert t_1/t_2 \rvert < 1$, and the red line represents the corresponding eigenvalues. In panel (f), we depict the quadrupole moment $Q_{xy}$ as a function of $t_1$. We choose the model parameters as $t_2=1.0$ [for all the panels] and $t_1=0.5$ [for the panels (a)-(d)]. 
	}
	\label{BBHresults}
\end{figure}

Afterward, we demonstrate a few numerical results related to the BBH model Hamiltonian that capture the HOTI phase. As the system represents an insulating phase, the bulk of the system exhibits a gapped band insulator. In Fig.~\ref{BBHresults}~(a), we depict the bulk bands for the Hamiltonian $\mathcal{H}_{\rm BBH} (\vect{k})$ as a function of the crystal momenta $k_x$ and $k_y$. One can observe a finite bulk gap separating the valence and the conduction bands. The edges should also be gapped in a system hosting a second-order phase. In Fig.~\ref{BBHresults} (b), we demonstrate the ribbon geometry (\ie the system is finite in one direction, say along $y$-direction and is periodic in the other direction) eigenvalue spectra $E$ of the BBH model Hamiltonian as a function of $k_x$. The edges also become massive in this case. However, the edges can be topological and support topological modes inside the edge gap. To obtain the features of the corner modes from the eigenvalue spectrum, we consider open boundary conditions (OBC) in both the $x$- and $y$-directions. We illustrate the eigenvalue spectrum $E_n$ (obtained via OBC) of the BBH model Hamiltonian as a function of the state index $n$ in Fig.~\ref{BBHresults} (c). One can clearly identify the presence of four zero-energy eigenvalues ($E=0$) from the inset of Fig.~\ref{BBHresults} (c). To demonstrate the localized nature of the corner modes, we compute the local density of states~(LDOS). In Fig.~\ref{BBHresults} (d), we depict the LDOS as a function of the system dimensions $x$ and $y$, computed at $E=0$. One can notice from the LDOS behavior that the zero-energy corner modes are sharply localized at the four corners of the system. To figure out the topological region for the Hamiltonian in the parameter space, we exhibit the eigenvalue spectra $E_n$ of the BBH model Hamiltonian obeying OBC as a function of the intra-cell hopping amplitude $t_1$ in Fig.~\ref{BBHresults} (e). It appears that the Hamiltonian $\mathcal{H}_{\rm BBH} (\vect{k})$ represents a 2D SOTI hosting localized zero-energy corner modes when the inter-cell hopping amplitude $t_2$ dominates over that of the intra-cell hopping \ie $\lvert t_2/t_1 \rvert > 1$. The red line represents the eigenvalues corresponding to the 0D corner states in Fig.~\ref{BBHresults} (e) when $\lvert t_2/t_1 \rvert > 1$.

The topological characterization for the SOTI phase can be achieved by computing the quadrupole moment $Q_{xy}$ with vanishing dipole moment for the bulk~\cite{benalcazar2017,benalcazarprb2017}. For a crystal obeying periodic boundary condition~(PBC), one may define the macroscopic quadrupole moment $Q_{xy}$ as~\cite{wheeler2018many,kang2018many}:
\begin{equation}\label{macroquadint}
	Q_{xy}=\frac{1}{2 \pi} \textrm{Im}\left[ \ln \bra{\Psi_0} e^{2 \pi i \sum_{r}\hat{q}_{xy}(\vect{r})}\ket{\Psi_0}\right]\ ,
\end{equation}
where, $\hat{q}_{xy}(\mathbf{r})=\frac{xy}{L^2}\hat{n}(\vect{r})$ represent the microscopic quadrupole moment at site $\vect{r}$, $L$ being the number of lattice sites considered along one direction, and $\ket{\Psi_0}$ is the many-body ground state which one can construct employing the occupied states~\cite{RestaPRL1998}. To compute the $Q_{xy}$ numerically, we first construct a $N\times N_{\rm occ}$ dimensional matrix $\mathcal{U}$ employing the sorted column-wise occupied eigenvectors of the real space BBH model Hamiltonian; with $N$ being the dimension of the Hamiltonian and $N_{\rm occ}$ represents the number of occupied eigenstates. Afterward, we formulate another matrix operator $\mathcal{W}^{Q}$ as
\begin{equation}\label{WforQuadint}
	\mathcal{W}_{ i\alpha,j}^{Q}= \exp \left[ {i \frac{2\pi}{L^2}f(x_{i\alpha},y_{i\alpha})} \right]  \: \mathcal{U}_{ i\alpha,j }\ ,
\end{equation}
Here, $\alpha$ represents all the orbital indices. Here,  $f(x_{i\alpha},y_{i\alpha})=x_{i\alpha}y_{i\alpha}$. Therefore, one can recast $Q_{xy}$ [Eq.~(\ref{macroquadint})] employing the $\mathcal{U}$ and $\mathcal{W}_{ i\alpha,j}^{Q}$ as  
\begin{equation}\label{Quadfinalint} 
	Q_{xy}=\frac{1}{2\pi} \;\textrm{Im} \;[\textrm{Tr}  \ln \left(\mathcal{U}^\dagger \mathcal{W}^{Q} \right)]\ .
\end{equation}
The quadrupole moment $Q_{xy}$ is, however, only defined up to modulo one, such that $Q_{xy}=Q_{xy}$ mod 1. We compute $Q_{xy}$ for the BBH model employing Eq.~(\ref{Quadfinalint}) and demonstrate it as a function of the intra-site hopping amplitude $t_1$ in Fig.~\ref{BBHresults} (f). The $Q_{xy}$ exhibit quantized value of $0.5$ in the topological regime \ie when $\lvert t_1 /t_2 \rvert < 1$. The mirror symmetries $\mathcal{M}_x$ and $\mathcal{M}_y$ play the pivotal role in the quantization of $Q_{xy}$~\cite{benalcazar2017,benalcazarprb2017}. Another important observation regarding the Hamiltonian  $\mathcal{H}_{\rm BBH} (\vect{k})$ is that, in $\mathcal{H}_{\rm BBH} (\vect{k})$, the terms associated with $k_x$ and $k_y$ are decoupled. Thus, one can recast $\mathcal{H}_{\rm BBH} (\vect{k})$ as two copies of the 1D Su-Schrieffer-Heeger (SSH) chain along two directions with specific matrix structure~\cite{LiPRL2020}.
The system is topological when both the 1D SSH chains represent a topological phase. 

\begin{figure}[]
	\centering
	\includegraphics[width=0.4\textwidth]{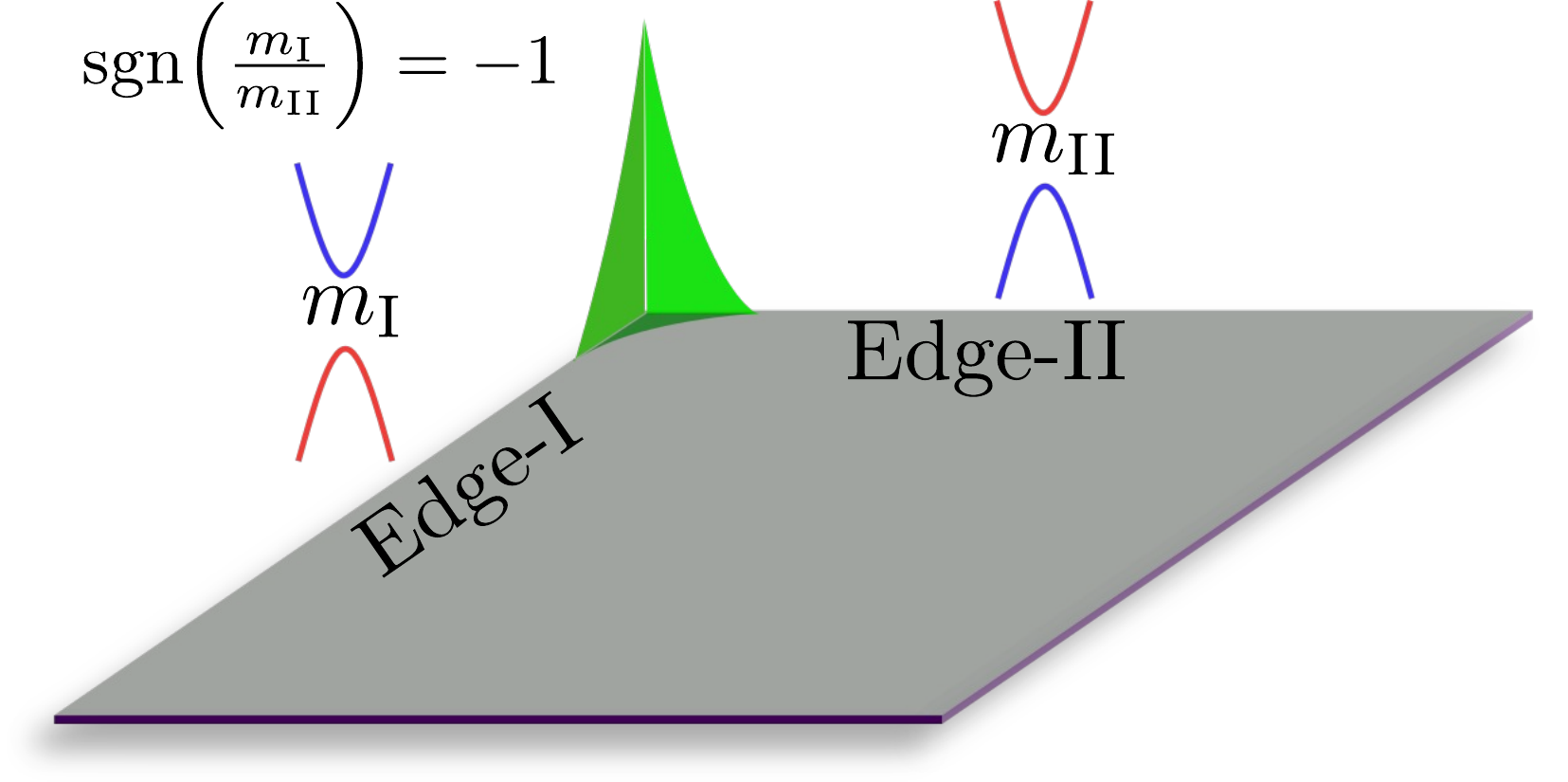}
	\caption{Schematic representation of the mass changing mechanism in a 2D SOTI across the intersecting edges to realize a zero-energy topological mode (represented by the green curve) at the corners of the system. The edge-I and -II mass gaps are assumed to be $m_{\rm I}$ and $m_{\rm II}$, respectively, such that sgn$\left(\frac{m_{\rm I}}{m_{\rm II}}\right)=-1$.
	}
	\label{JWSOTI}
\end{figure}

The appearance of the 0D corner modes can be understood analytically by constructing the Hamiltonians for the edges while starting from the bulk Hamiltonian~\cite{benalcazar2017,SchindlerDirac2020}. In a 2D SOTI, the 1D edges can be represented using 
Dirac Hamiltonians with mass terms. We schematically demonstrate this scenario in Fig.~\ref{JWSOTI}. Two intersecting edge-Hamiltonians corresponding to the edge-I and -II carry mass terms $m_{\rm I}$ and $m_{\rm II}$, respectively. In the topological phase, these two masses carry opposite signs such that sgn $\left(\frac{m_{\rm I}}{m_{\rm II}}\right)=-1$. Afterward, the knowledge of Jackiw-Rebbi theorem~\cite{jackiw1976solitons} enables us to identify the change of sign of the mass terms of the edge Hamiltonians intersecting at a corner, and thus one can find the solution for the zero-energy localized corner state~\cite{benalcazar2017,SchindlerDirac2020}. 

\subsubsection{Model-2: The BHZ model with $C_4$ and $\mathcal{T}$ breaking Wilson-Dirac mass term}
The 2D SOTI phase can also be realized considering the BHZ model \cite{bernevig2006quantum} with 
a WD mass term~\cite{schindler2018,Roy2019}. In contrast to the BBH model where one can only realize the SOTI phase, this model allows us to study the hierarchy of topological orders. The Hamiltonian describing this system can be written as
\begin{align}
	\mathcal{H}_{\rm BHZ+WD} (\vect{k}) =& \lambda \sin k_x  \ \sigma_x s_z +  \lambda \sin k_y  \ \sigma_y s_0 \non \\
	&+ \left( m_0 -  t \cos k_x - t \cos k_y \right)  \sigma_z s_0 \non \\
	&+ \Lambda \left(\cos k_x -\cos k_y \right) \sigma_x s_x \ ,
	\label{BHZ_C4}
\end{align}
\begin{figure}[]
	\centering
	\includegraphics[width=0.49\textwidth]{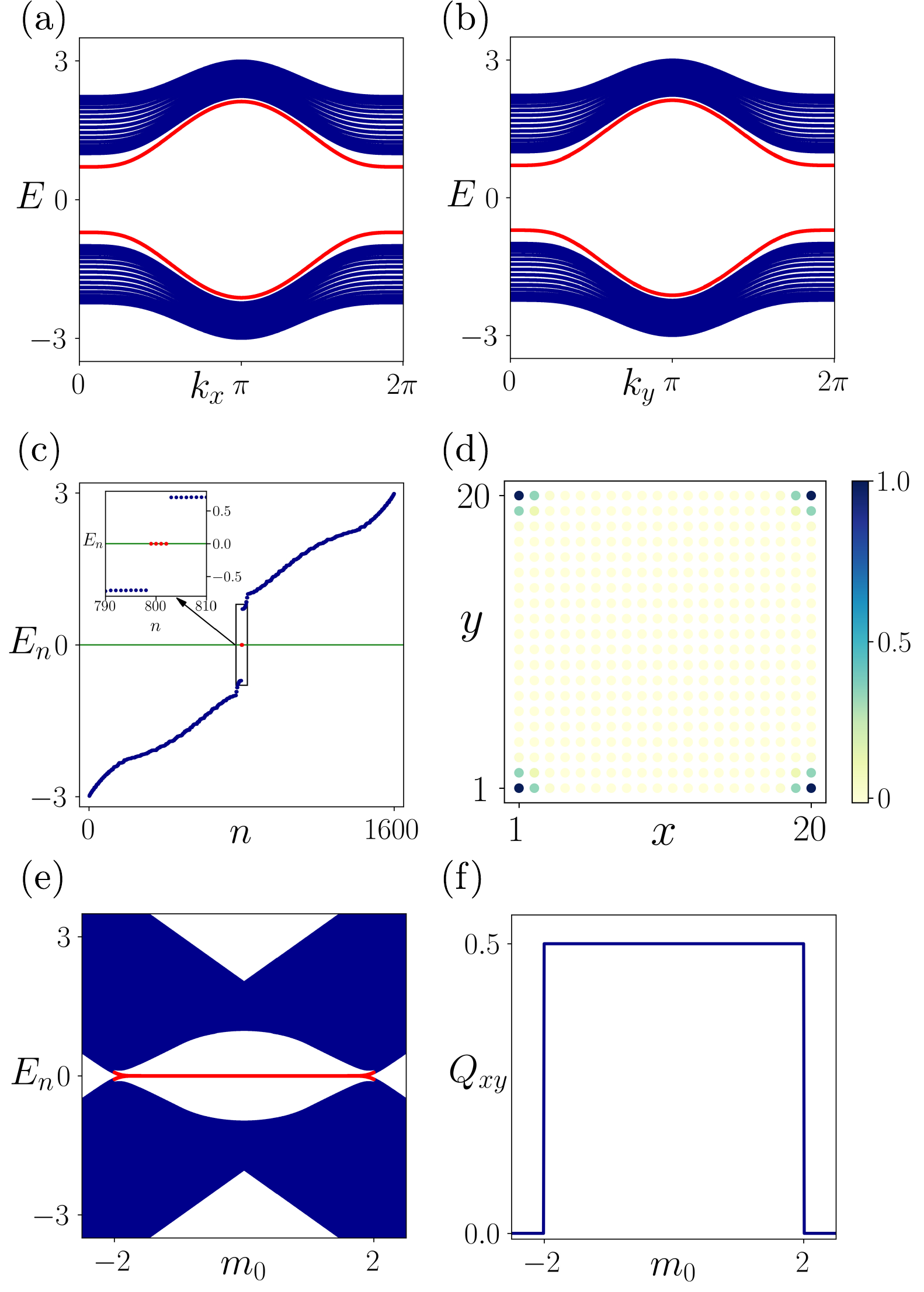}
	\caption{In panels (a) and (b), we depict the gapped spectrum considering the ribbon geometry 
for the Hamiltonian [Eq. (\ref{BHZ_C4})], as a function of the momenta $k_x$ and $k_y$, respectively. 
In panel (c), we depict the eigenvalue spectrum as a function of the state index $n$ for the finite size system ($20 \times 20$ lattice sites). The LDOS associated with the zero-energy corner states is demonstrated in panel (d). In panel (e), we show the eigenvalue spectra of the finite-size system as a function of $m_0$. In panel (f), we depict the quadrupole moment $Q_{xy}$ as a function of $m_0$. We choose the model parameters as follows: $\lambda=t=m_0=\Lambda=1$ [for the panels (a)-(d)].
	}
	\label{BHZresults}
\end{figure}
\hspace{-0.15cm}where, $t$ represents the nearest neighbor hopping amplitude, $\lambda$ denotes the SOC strengh, $m_0$ symbolize the crystal field splitting, and $\Lambda$ indicates the amplitude of the WD mass term. The Pauli matrices $\vect{\sigma}$ and $\vect{\tau}$ act on the orbital and the spin degrees of freedom, respectively. Below, we discuss various symmetry properties of the Hamiltonian $\mathcal{H}_{\rm BHZ+WD} (\vect{k})$: 
\begin{itemize}
	\item TRS with $\mathcal{T}= i \sigma_0 s_y\mathcal{K}$: $\mathcal{T} \mathcal{H}(\vect{k}) \mathcal{T}^{-1}=\mathcal{H}(-\vect{k})$, if $\Lambda=0$,
	\item Charge conjugation symmetry with $\mathcal{C}=\sigma_x s_z\mathcal{K}$:  $\mathcal{C} \mathcal{H}(\vect{k}) \mathcal{C}^{-1}= -\mathcal{H}(-\vect{k})$,
	\item Sublattice or chiral symmetry with $\mathcal{S}=\sigma_x s_y\mathcal{K}$:  $\mathcal{S} \mathcal{H}(\vect{k}) \mathcal{S}^{-1}= -\mathcal{H}(\vect{k}) $,
	\item Four-fold rotation symmetry with $C_4=  e^{-i \frac{\pi}{4} \sigma_z s_z}$: $C_4 \mathcal{H}(k_x,k_y)C_4^{-1} = \mathcal{H}(-k_y,k_x)$, if $\Lambda = 0$, 
	\item The combined $C_4 \mathcal{T}$ symmetry: $(C_4 \mathcal{T}) \mathcal{H}(k_x,k_y)(C_4 \mathcal{T})^{-1} = \mathcal{H}(k_y,-k_x)$.
\end{itemize}
The difference between Model-1 and Model-2 is that the former can only hosts second-order phase while the latter can host first- as well as second-order phase both. Specifically, when $\Lambda$ 
is zero, the Hamiltonian [Eq.~(\ref{BHZ_C4})] represents a QSHI hosting gapless 1D edge-modes if $-2t<m_0 < 2t$. However, adding the WD mass term introduces a gap in the edges. The edge-Hamilonians corresponding to the two intersecting 1D edges inherit opposite masses, proportional to $\Lambda$. The Jackiw-Rebbi theorem then guarantees to perceive the emergence of the 0D 
corner modes. 

Having a phenomenological understanding of this model, we tie up with a few numerical results related to this system. The bulk bands of $\mathcal{H}_{\rm BHZ+WD} (\vect{k})$ exhibit a finite gap around 
the Fermi energy resembling that of the BBH model [see Fig.~\ref{BBHresults}~(a)]. In Figs.~\ref{BHZresults} (a) and (b), we depict the ribbon geometry spectrum with a finite number of lattice sites along $y$ and $x$, respectively. It is evident that a finite gap exists in both the edge states. Nevertheless, the eigenvalue spectrum of the finite-size system Hamiltonian should manifest zero-energy eigenvalues if the corner states are present. In Fig.~\ref{BHZresults} (c), we depict the eigenvalue spectrum $E_n$ as a function of the state index $n$ for the Hamiltonian obeying OBC along both $x$ and $y$ directions. The zero-energy states are denoted by the red points in the inset of Fig.~\ref{BHZresults} (c). To ascertain the corner localization of the zero-modes, we compute the LDOS and represent the same as a function of the system dimensions $x$ and $y$ in Fig.~\ref{BHZresults} (d). It is apparent that the zero-energy states are sharply localized at the corners of the system. Thus, one can confirm that the Hamiltonian $\mathcal{H}_{\rm BHZ+WD} (\vect{k})$ possesses the characteristic of a 2D SOTI. In Fig.~\ref{BHZresults} (e), we illustrate the eigenvalue spectra of the Hamiltonian as a function of the crystal field splitting mass term $m_0$, obeying OBC in both the directions $x$ and $y$. The SOTI phase is obtained when $-2t<m_0 < 2t$. Furthermore, we compute the quadrupole moment $Q_{xy}$ for this model [see Eq.~(\ref{Quadfinalint})] and demonstrate the same as a function of $m_0$ in Fig.~\ref{BHZresults} (f). We obtain $Q_{xy}=0.5$ mod 1, for $-2t<m_0 < 2t$ and $0$, otherwise. This confirms the second-order band topology of this model.

\subsection{3D SOTI}
\begin{figure}[]
	\centering
	\includegraphics[width=0.49\textwidth]{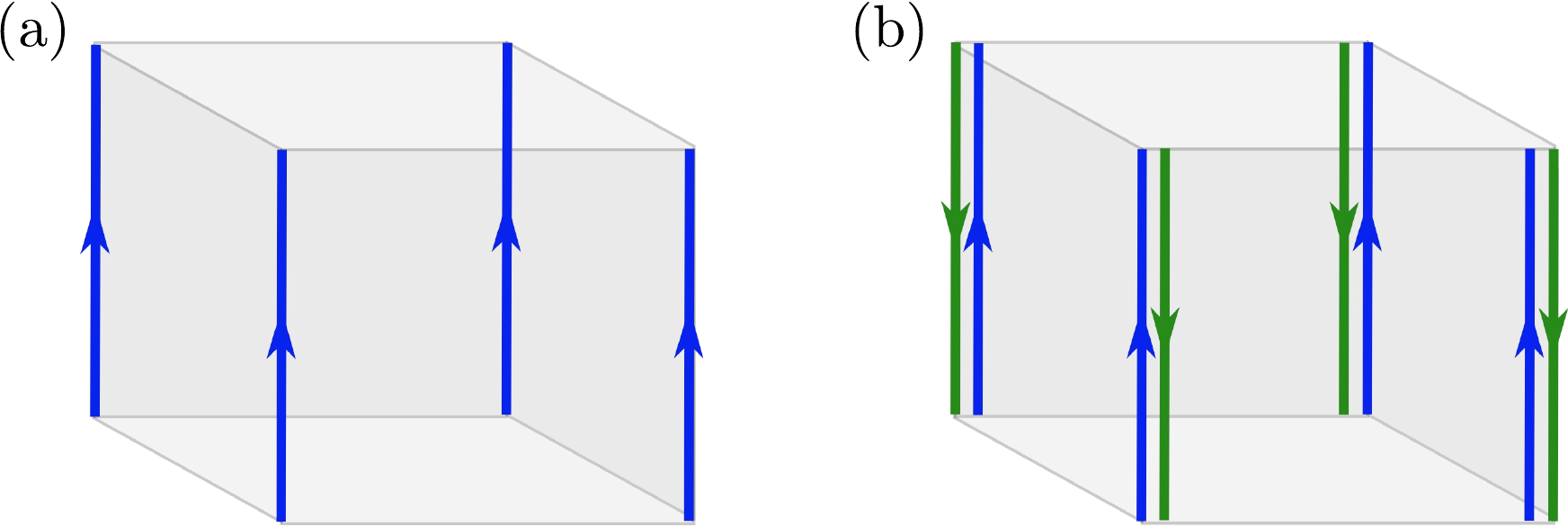}
	\caption{Schematic diagram of a 3D SOTI hosting chiral and helical hinge modes are presented 
in panels (a) and (b), respectively. The arrows on the blue and green lines indicate the propagation direction of the electrons along the hinges.
	}
	\label{ChiralHelicalSOTI}
\end{figure}

In three dimensions, the SOTIs are characterized by a gapped bulk and surface state while exhibiting gapless hinge states~[see Fig.~\ref{static} (c)]. Unlike a 2D SOTI, a 3D SOTI exhibits gapless dispersive modes which appear at the hinges of the system. These hinge modes can either be chiral or helical [see Fig.~\ref{ChiralHelicalSOTI}]. The chiral SOTI breaks TRS, and thus the chiral hinge modes are unidirectional [see Fig.~\ref{ChiralHelicalSOTI} (a)] akin to the quantum Hall or QAH insulator edge states. In contrast, the helical SOTI respects TRS, and thus the helical hinge modes are accompanied 
by a counter-propagating partner [see Fig.~\ref{ChiralHelicalSOTI} (b)] as guranteed by the Kramers' degeneracy resembling the QSHI~\cite{schindler2018}.

\begin{figure}[]
	\centering
	\includegraphics[width=0.49\textwidth]{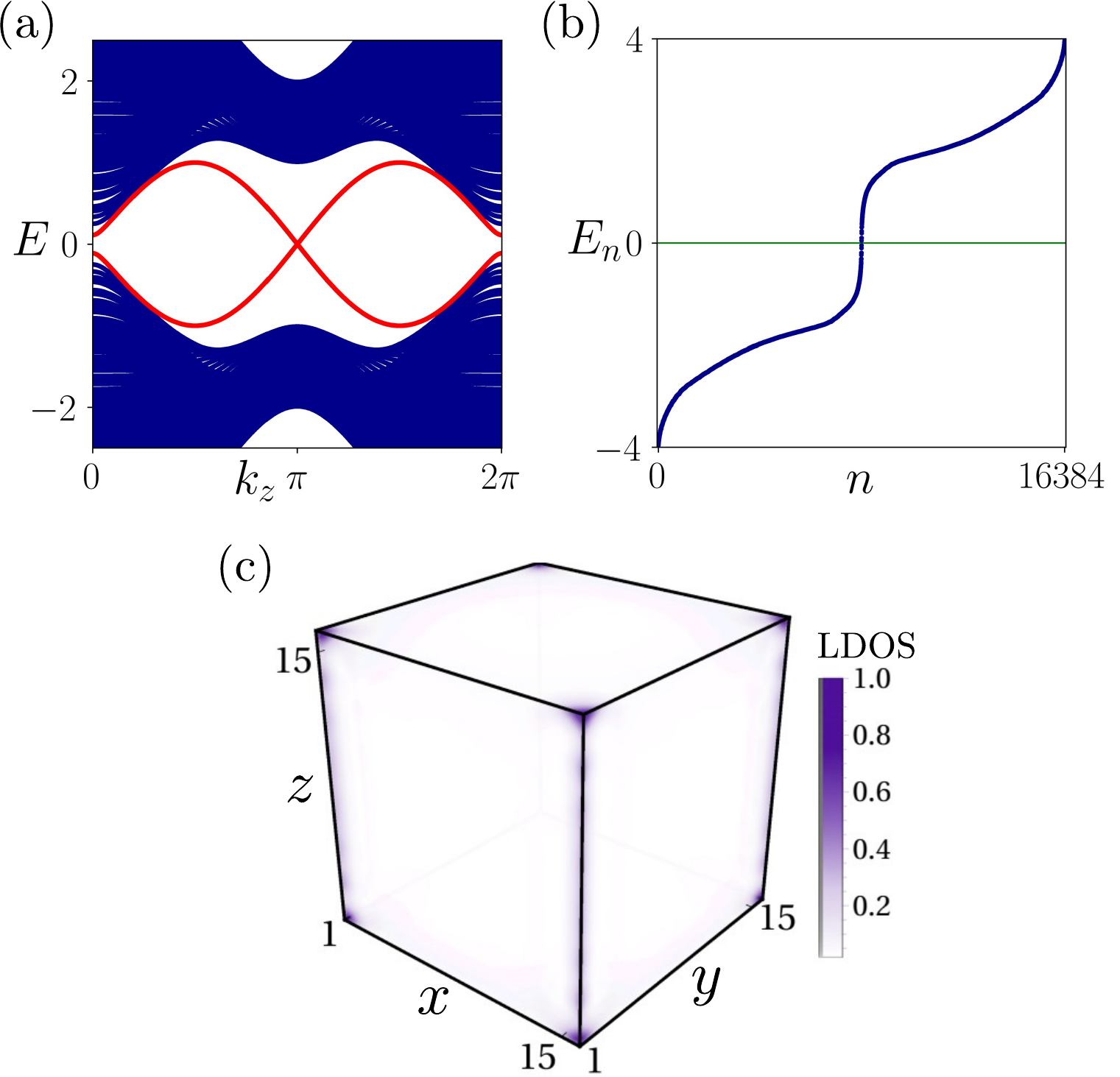}
	\caption{The eigenvalue spectrum corresponding to the rod geometry Hamiltonian of a 3D SOTI is depicted in panel (a) as a function of $k_z$. In panel (b), we demonstrate the eigenvalue spectrum $E_n$ of the finite size ($15 \times 15 \times 15$ cubic lattice) Hamiltonian as a function of the state index $n$. The LDOS distribution associated with the states near $E_n=0$ is shown in panel (c) as a function of the system dimensions. We choose the model parameters as follows: $\lambda=t=m_0=\Lambda=1$.
	}
	\label{3DSOTIresult}
\end{figure}

Here, we explore the model of a 3D SOTI proposed by Schindler \etal~\cite{schindler2018}. The corresponding model Hamiltonian reads as
\begin{align}
	\mathcal{H}_{\rm SOTI}^{\rm 3D} (\vect{k}) =& \lambda \sin k_x  \ \sigma_x s_x +  \lambda \sin k_y  \ \sigma_x s_y + \lambda \sin k_z  \ \sigma_x s_z \non\\
	&+  \left( m_0 +  t \sum_{i=x,y,z}  \cos k_i \right)  \sigma_z s_0 \non \\
	&+ \Lambda \left(\cos k_x -\cos k_y \right) \sigma_y s_0 \ ,
	\label{3D_SOTI}
\end{align}
where, the Pauli matrices $\vect{\sigma}$ and $\vect{s}$ operate on the orbital and spin spaces, respectively. Here, $t$, $\lambda$, $m_0$, and $\Lambda$ represent the amplitude of the hopping, SOC, crystal field splitting, and the $C_4$ and TRS braking WD mass term, respectively. Similar to the 2D SOTI Hamiltonian, this Hamiltonian [$\mathcal{H}_{\rm SOTI}^{\rm 3D} (\vect{k})$] breaks both TRS $\mathcal{T}$ and four-fold rotation symmetry $C_4$ when $\Lambda \neq 0$; with $\mathcal{T}=i \sigma_0 s_y \mathcal{K}$ and $C_4=\sigma_0 e^{-\frac{\pi}{4}s_z}$. Nevertheless, $\mathcal{H}_{\rm SOTI}^{\rm 3D}(\vect{k})$ respects the combined $C_4 \mathcal{T}$ symmetry: $(C_4 \mathcal{T}) \mathcal{H}_{\rm SOTI} (k_x, k_y, k_z) (C_4 \mathcal{T})^{-1}= \mathcal{H}_{\rm SOTI} (k_y, -k_x, -k_z)$. Moreover, when $\Lambda=0$ the Hamiltonian [Eq.~(\ref{3D_SOTI})] exhibits a strong 3D TI phase with 2D gapless surface states if $t < \lvert m_0 \rvert < 3$~\cite{hasan2010colloquium,FuKane2007,Zhang2009}. When $\Lambda \neq 0$, the surface states are gapped out by a mass term proportional to $\Lambda$. However, the $yz$- and $zx$-surface accumulate mass terms that are opposite in sign. Thus employing the Jackiw-Rebbi theorem, one can obtain a hinge mode along the junction of the above-mentioned surfaces~\cite {Song2017,jackiw1976solitons}.

To identify the hinge states, we employ rod geometry \ie the system obeys PBC along one direction, say $z$-direction, and satisfies OBC along the remaining two directions \ie $x$- and $y$-directions. In the rod geometry, we diagonalize the Hamiltonian and demonstrate the eigenvalues as a function of the momentum $k_z$ in Fig.~\ref{3DSOTIresult} (a). Here, red lines represent the gapless dispersive nature of the hinge modes. Moreover, in a finite size system \ie OBC along all three directions, the eigenvalue spectrum $E_n$ of the Hamiltonian manifest a gapless nature around the energy $E_n=0$ when illustrated as a function of the state index $n$ [see Fig.~\ref{3DSOTIresult} (b)]. The LDOS distribution corresponding to the zero-energy states is shown in Fig.~\ref{3DSOTIresult} (c) as a function of the system dimensions $x$, $y$, and $z$. It is evident that the hinges modes are sharply localized along the hinges while decaying exponentially into the surfaces as well as in the bulk. The corresponding decaying/localization length of the hinge modes depends on the system parameters \ie $\lambda, t, m_0, \Lambda$ etc.

\subsection{3D TOTI} \label{Sec:3DTOTIstatic}
\begin{figure}[]
	\centering
	\includegraphics[width=0.47\textwidth]{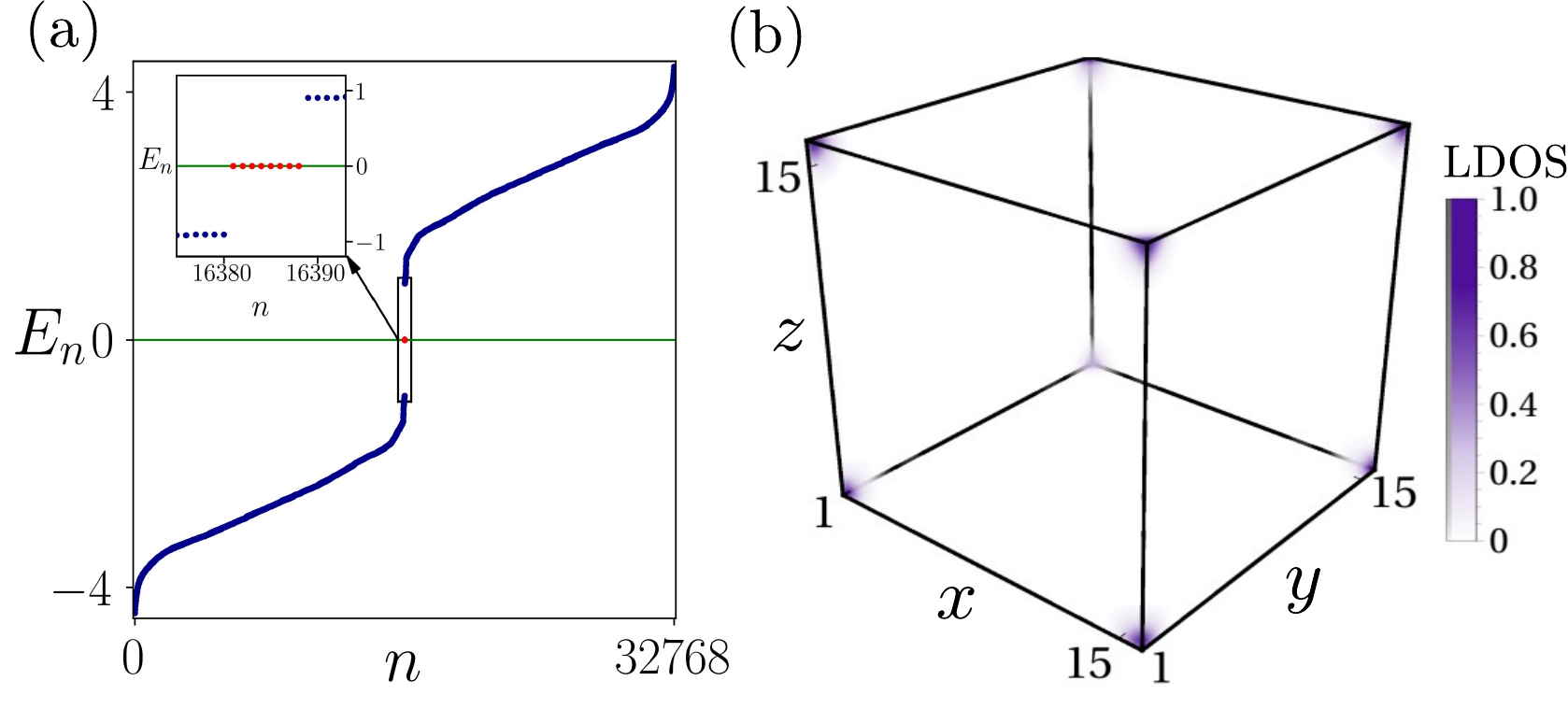}
	\caption{(a) Eigenvalue spectrum $E_n$ is depicted as a function of the state index $n$ for the finite size Hamiltonian [$15 \times 15 \times15$ cubic lattice] of 3D TOTI, obeying OBC along all directions. In the inset, the eigenvalues close to $E_n=0$ are highlighted for better clarity. (b) The LDOS distribution associated with the $E_n=0$ states is illustrated as a function of the system dimension $x$, $y$, and $z$. This confirms the sharp corner localization of the zero-energy modes. We choose the model parameters as follows: $\lambda=t=m_0=\Lambda_1=\Lambda_2=1$.
	}
	\label{3DTOTIresult}
\end{figure}

In three dimensions, one can also realize a third-order topological phase apart from the first and second-order phases. The 3D TOTIs exhibit gapped bulk, surface, and hinge states [see Fig.~\ref{static} (e)]. Nevertheless, a TOTI manifests localized 0D corner modes akin to the 2D SOTI. A mass term gaps out the hinge modes of a TOTI such that two adjacent hinges meeting at a corner accommodate mass terms of opposite signs. Afterward, the Jackiw-Rebbi theorem can be utilized to understand the appearance of the corner modes. The 3D TOTI can be thought of as an octupolor system~\cite{benalcazar2017,benalcazarprb2017}. To host an octupolar phase (3D TOTI) one needs at least four unoccupied bands~\cite{benalcazar2017,benalcazarprb2017}. A 3D version of the BBH model can 
be employed to realize only the 3D TOTI if one can couple three 1D SSH chains along three orthogonal directions with appropriate matrix structure. However, the BBH model does not allow us to systematically study the appearance of different orders of topological phases as discussed earlier. On the other hand, one can start with the Hamiltonian of a 3D SOTI [Eq.~(\ref{3D_SOTI})] and introduce another degree of freedom (sublattice). Afterward, the hinge modes can be gapped out by incorporating another appropriate WD mass term~\cite{Nag2021}. Below we present the model Hamiltonian that represents a 3D TOTI.
\begin{widetext}
\vspace{-1.0cm}
	\begin{align}
		\mathcal{H}_{\rm TOTI}^{\rm 3D} (\vect{k}) =& \lambda \sin k_x  \ \mu_x \sigma_x s_x +  \lambda \sin k_y  \ \mu_x \sigma_x s_y + \lambda \sin k_z  \ \mu_x \sigma_x s_z  +  \left( m_0 + t \sum_{i=x,y,z}  \cos k_i \right) \mu_x \sigma_z s_0 \non \\
		&+  \Lambda_1 \left(\cos k_x -\cos k_y \right) \mu_x \sigma_y s_0  + \Lambda_2 \left(2 \cos k_z -\cos k_x -\cos k_y \right) \mu_z \sigma_0 s_0 \ ,
		\label{3D_TOTI}
	\end{align}
\end{widetext}
where, the Pauli matrices $\vect{\mu}$, $\vect{\sigma}$, and $\vect{s}$ act on the sublattice, orbital, and spin degrees of freedom, respectively. Here, $t$, $\lambda$, $m_0$, and $\Lambda_{1,2}$ represent the amplitude of the hopping, SOC, crystal field splitting, and WD mass terms, respectively. We demonstrate the eigenvalue spectrum $E_n$ corresponding to the Hamiltonian [Eq.~(\ref{3D_TOTI})] as a function of the state index $n$ in Fig.~\ref{3DTOTIresult} (a). Here, the system obeys OBC in all three directions ($x$, $y$, and $z$). The eight zero-energy eigenvalues are marked by the red points for clarity in the inset of Fig.~\ref{3DTOTIresult} (a). The LDOS at $E_n=0$ is illustrated in Fig.~\ref{3DTOTIresult} (b). The zero-energy states are clearly localized at the eight corners of the system. When $\Lambda_1 \neq 0$ and $\Lambda_2=0$, the system exhibits a 3D SOTI hosting 1D propagating hinge modes. Thus, this model enables one to investigate the hierarchy of different higher-order topological phases systematically.

The 3D TOTI phase can be topologically characterized employing the octupole moment $O_{xyz}$. The macroscopic octupole moment $O_{xyz}$ for a crystal obeying PBC can be defined through the microscopic octupolar moment operator $\hat{o}_{xyz}(\vect{r})=\frac{xyz}{L^3}\hat{n}(\vect{r})$ as~\cite{wheeler2018many,kang2018many} :
\begin{equation}\label{macrooctint}
	O_{xyz}=\frac{1}{2 \pi} \textrm{Im}\left[ \ln \bra{\Psi_0} e^{2 \pi i \sum_{r}\hat{o}_{xyz}(\vect{r})}\ket{\Psi_0}\right]\ .
\end{equation}
Here, $\ket{\Psi_0}$ is the many-body ground state. As before, we first construct a $N\times N_{\rm occ}$ dimensional matrix $\mathcal{U}$ employing the $N_{\rm occ}$ eigenvectors corresponding to the real space model Hamiltonian of TOTI. Afterward, we formulate another matrix operator $\mathcal{W}^O$ as
\begin{equation}\label{WforOctint}
	\mathcal{W}_{ i\alpha,j}^O = \exp \left[ {i \frac{2\pi}{L^2}f(x_{i\alpha},y_{i\alpha},z_{i\alpha})} \right]  \: \mathcal{U}_{ i\alpha,j }\ .
\end{equation}
where, $\alpha$ represents the sublattice, orbital, and spin indices. Here, $f(x_{i\alpha},y_{i\alpha},z_{i\alpha})=x_{i\alpha}y_{i\alpha}z_{i\alpha}$. Therefore, $O_{xy}$, defined in Eq.~(\ref{macrooctint}), 
can be recast in the form 
\begin{equation}\label{Octfinalint}
	O_{xyz}=\frac{1}{2\pi} \;\textrm{Im} \;[\textrm{Tr}  \ln \left(\mathcal{U}^\dagger \mathcal{W} \right)]\ .
\end{equation}
In the topological regime a TOTI exhibits a quantized $O_{xyz}=0.5~({\rm mod}~1)$~\cite{benalcazar2017,benalcazarprb2017}, when both $\Lambda_{1,2} \neq 0$.

\section{Floquet Theory and Floquet FOTI: A primer}  \label{sec3}
Here, we briefly discuss the main outline of Floquet theory and its application for the driven BHZ model. The latter turns out to be a Floquet FOTI~(FFOTI). The key parameters in a driven system are the band-width in terms of hopping amplitude ($\sim$~eV), SOC ($\sim$~meV) and the driving frequency ($\sim$~THz).
\subsection{Floquet theory}
In a periodically driven system, the Hamiltonian governing the system can be written in the form
\begin{equation}
	H(t)=H(t+T)=H_0+V(t) \ ,
	\label{periodicHam}
\end{equation}
where, $T=(2 \pi / \omega)$ is the time-period of the drive. Here, $H_0$ and $V(t)$ represent the static and the time-dependent part of the full Hamiltonian $H(t)$ with $V(t)$ satisfying $V(t)=V(t+T)$. The static Hamiltonian $H_0$ is assumed to have the eigenvalues $\left\{E_n\right\}$ and eigenfunctions $\left\{\ket{\phi_n}\right\}$. The time-dependent Schr{\"o}dinger equation for $H(t)$ reads as
\begin{equation}
	\left[H(t) - i \hbar \frac{\partial}{\partial t} \right] \ket{\Psi (t)} =0 \ .
	\label{time_sch}
\end{equation}
Now, employing the Floquet theorem~\cite{FLoquetpaper1883,GrifoniDrivenquantum1998}, one obtains the solution of Eq.~(\ref{time_sch}) as
\begin{equation}
	\ket{\Psi_\alpha(t)}=e^{-i \epsilon_\alpha t /\hbar}  \ \ket{\Phi_\alpha (t)}    \ ,
	\label{FLoquet_states}
\end{equation}
where, $\ket{\Psi_\alpha(t)}$'s are called the Floquet states~\cite{ShirleyPRB1965,SambeFloquet1973}, while $\ket{\Phi_\alpha (t)}$'s are the time-periodic Floquet modes such that: $\ket{\Phi_\alpha (t+T)}=\ket{\Phi_\alpha (t)}$. The $\epsilon_\alpha$ is the quasienergy analogous to the Bloch momentum in a periodic solid. The Floquet states are also the eigenstates of the time-evolution operator over one driving period $T$ (also known as the Floquet operator), such that
\begin{equation}
	U(t_0+T,t_0) \ket{\Psi_\alpha(t_0)} = e^{-i \epsilon_\alpha t /\hbar}  \ket{\Psi_\alpha(t_0)} \ ,
	\label{floquetop}
\end{equation}
where $t_0$ indicates the initial time. However, the quasienergy $\epsilon_\alpha$ is independent of the choice of $t_0$~\cite{Eckardt2015}. Time-dependent Floquet states at any time instance $t$ can be obtained by operating the time-evolution operator as
\begin{equation}
	\ket{\Psi_\alpha(t)}=U(t,t_0)\ket{\Psi_\alpha(t_0)} \ .
\end{equation}
On the other hand, the Floquet operator $U(t_0+T,t_0)$ can also be computed in terms of a time-ordered~(TO) notation as follows 
\begin{align}
	U(t_0+T,t_0)=& {\rm TO} ~ \exp \left[ -i \int_{t_0}^{t_0+T} dt H(t) \right]  \non \\
	=& \prod_{j=0}^{N-1} U(t_{j}+\delta t,t_j) \ ,
	\label{FloquetOptime}
\end{align}
where, $U(t_{j}+\delta t,t_j)=e^{-i H(t_j)\delta t}$; with $\delta t = \frac{T}{N}$ and $t_j=t_0+j \delta t$. However, $U(t_{j}+\delta t,t_j)$ can be computed more efficiently employing the second-order Trotter-Suzuki formalism~\cite{DAlessio2015,Suzuki1976,DeRaedt183,QinPRB2022} as follows
\begin{eqnarray}\label{TS_time-evoCh2}
	U(t_{j}+\delta t,t_j)= e^{- i  \frac{\delta t}{2} V\left(t_j + \frac{\delta t}{2} \right)}  e^{-i \delta t H_0}  e^{- i  \frac{\delta t}{2} V\left(t_j + \frac{\delta t}{2} \right)}  . \ \ \
\end{eqnarray} 
Note that, one needs to evaluate $e^{-i \delta t H_0}$ only once as it is independent of time. However, the time increment $\delta t$ needs to be chosen in such a way that $U(T,0)$ is unitary. Following \Eq{FloquetOptime}, one can also construct a time-evolution operator at any time $t$, $U(t+t_0,t_0)$ by replacing $T \rightarrow t$. 

On the other hand, employing the Floquet operator $U(t_0+T,t_0)$, one can also define a time-independent Floquet Hamiltonian $H_{\rm F}$, such that
\begin{equation}
	U(t_0+T,t_0)=\exp \left(-\frac{i H_{\rm F} T}{\hbar}\right) \ .
\end{equation}
The Floquet Hamiltonian $H_{\rm F}$ shares the same Floquet states $\ket{\Psi_\alpha(t_0)}$ as the Floquet operator and one can represent $H_{\rm F}$ in terms of the Floquet modes $\ket{\Phi_\alpha(t_0)}$ as
\begin{equation}
	H_{\rm F}=\sum_\alpha \epsilon_\alpha \ket{\Psi_\alpha(t_0)} \bra{\Psi_\alpha(t_0)}  \ .
	\label{FloquetHam1}
\end{equation}
The Floquet Hamiltonian $H_{\rm F}$ along with the Floquet operator $U(t_0+T,t_0)$ serves the purpose of the dynamical analog of the static Hamiltonian.

At first glance, one can identify that the phase-factor $e^{-i \epsilon_\alpha t /\hbar}$ in Eq.~(\ref{floquetop}) is not uniquely defined. One may replace $\epsilon_\alpha$ by $\epsilon_{\alpha m}=\epsilon_\alpha+m \omega$ with $m \in \mathbb{Z}$ and the exponential still remain invariant. Thus, there is an ambiguity in defining the Floquet modes $\ket{\Phi_\alpha(t)}=e^{i \epsilon_\alpha t /\hbar}  \ \ket{\Psi_\alpha (t)} $ as well the Floquet Hamiltonian $H_{\rm F}$ in Eq.~(\ref{FloquetHam1}). However, one can employ the idea of the Brillouin zone used in a periodic lattice to restrict the quasi-momenta in the first Brilloun-zone. Here, one may as well invoke an analogous first Floquet zone such that $\epsilon_\alpha$ is defined using a modulo operation such that $\epsilon_{\alpha}= \epsilon_{\alpha m}~{\rm mod}~\hbar \omega$. In the extended space of quasienergies, the Floquet modes take the form
\begin{equation}
	\ket{\Phi_{\alpha m}(t)}=e^{i \epsilon_{\alpha m} t /\hbar}  \ \ket{\Phi_\alpha (t)} \ ,
\end{equation}
so that 
\begin{equation}
	\ket{\Psi_{\alpha}(t)}=e^{-i \epsilon_\alpha t /\hbar}  \ket{\Phi_\alpha (t)} =e^{-i \epsilon_{\alpha m} t /\hbar}  \ \ket{\Phi_{\alpha m} (t)} \ .
\end{equation}
Now, the Floquet modes $\ket{\Phi_{\alpha}(t)}$ can be expressed in terms of Fourier components as
\begin{equation}
	\ket{\Phi_{\alpha}(t)}= \sum_{m \in \mathbb{Z}} e^{-i m \omega t} \ket{\varphi_{\alpha}^{(m)}} \ ,
\end{equation}
where, $\ket{\varphi_{\alpha}^{(m)}}$'s are the Fourier component of $\ket{\Phi_{\alpha}(t)}$. In terms of these Fourier components, the Floquet states $\ket{\Psi_{\alpha}(t)}$ [\Eq{FLoquet_states}] reads
\begin{equation}
	\ket{\Psi_{\alpha}(t)}= \sum_m e^{-i \left(\epsilon_\alpha+m \omega \right) t / \hbar} \ket{\varphi_{\alpha}^{(m)}} \ .
	\label{floquetstatesFourier}
\end{equation}
After substituting Eq.~(\ref{floquetstatesFourier}) in Eq.~(\ref{time_sch}), we obtain
\begin{equation}
	\sum_{m' \in \mathbb{Z}} \left[ H_{m-m'} - m \omega \delta_{m , m'}  \right] \ket{\varphi_{\alpha}^{(m')}} = \epsilon_\alpha \ket{\varphi_{\alpha}^{(m)}} \ ,
	\label{Floqurtmodeseq}
\end{equation}
where we have introduced the Fourier component of $H(t)$ as 
\begin{equation}\label{FourierCompH} 
	H_{m} = \int_{t_0}^{t_0+T} \frac{d t}{T} H(t) \ e^{i m \omega t} \ .
\end{equation}
Here, the original Hilbert space $\mathbb{H}$ of $H(t)$ is expanded to $\mathbb{H} \otimes \mathbb{T}$~\cite{MikamiBW2016,ShirleyPRB1965,SambeFloquet1973}. The Hilbert space $\mathbb{T}=\oplus_m \mathbb{T}_m$ is spanned by the time-periodic functions such that $\mathbb{T}_m=\left\{ e^{-i m \omega t}\right\}$ and $m \in \mathbb{Z}$. Afterward, we can construct the infinite-dimensional time-independent Floquet Hamiltonian, incorporating the frequency-zone scheme as~\cite{Eckardt2015,ShirleyPRB1965,SambeFloquet1973}
\begin{widetext}
	\begin{equation}\label{FloquetHam}
		H^\infty_{\rm F}=
		\begin{pmatrix}
			\ddots& & &  \vdots & & & \udots\\
			&H_{-2} & H_{-1} & H_0  -2 \hbar \omega &  H_{1} &  H_{2} \\
			& & H_{-2} & H_{-1} & H_0 -\hbar \omega &  H_{1} &  H_{2} &    \\
			& & & H_{-2} & H_{-1} & H_0 & H_{1} & H_{2}\\
			& & &  & H_{-2} & H_{-1} & H_0 + \hbar \omega &  H_{1} &  H_{2}  \\
			& & & & & H_{-2} & H_{-1} & H_0 +2 \hbar \omega &  H_{1} &  H_{2} \\
			&&  & &  \udots & & &\vdots  & & \hspace{1cm}\ddots\\
		\end{pmatrix} \ .
	\end{equation}
\end{widetext}
Here, the structure of $H^\infty_{\rm F}$ is analogous to a quantum system coupled to a photon-like bath with $m$ being the photon number~\cite{Eckardt2015}. The $H_m$'s $(\lvert m \rvert \ge 1)$ describes a $m$-photon process and acts as the coupling between different photon sectors. Although, it is a formidable task to deal with this infinite dimensional Hamiltonian $H^\infty_{\rm F}$. Nevertheless, in the high-frequency limit \ie $\omega \gg$ band-width of $H_0$, one may employ perturbation theories like Brillouin-Wigner~(BW) perturbation theory~\cite{MikamiBW2016}, van Vleck perturbation theory~\cite{Eckardt2015,MikamiBW2016}, Floquet-Magnus perturbation theory~\cite{casas2001floquet,blanes2009magnus,Eckardt2015,MikamiBW2016}, etc. However, in the moderate frequency regime \ie $\omega \sim$ bandwidth, such perturbation theory breaks down. Nevertheless, one may consider a few photon sectors by defining a suitable cutoff~\cite{Piskunow2014,Usaj2014}. However, in the low-frequency regime, there is some development in the direction of deriving an effective Hamiltonian~\cite{Rodriguez-Vega2018,MichaelPRB2020}. On the other hand, one may resort to high-amplitude perturbation theory or Floquet perturbation theory when the amplitude of the driving field is much larger than the energy scales of the static Hamiltonian~\cite{Sen2021,Mukherjee2020,Sengupta2022Weyl,GhoshFPT2023}. Another approach to obtain an effective Hamiltonian description of the driven system is via constructing a $(d+1)$-dimensional time-independent Hamiltonian starting from a $d$-dimensional time-periodic system~\cite{PlateroPRL2013}.

\subsection{Floquet FOTI: Driven BHZ model }
The topological transition in the BHZ model considering the HgTe/CdTe quantum well can be tuned by the critical thickness of the quantum well~\cite{bernevig2006quantum,konig2007quantum}. 
This restricts the possible ways to tune the topological phases of the matter due to the unavailability of suitable knobs to control the band topology~\cite{NHLindner2020}. However, one can add a time-periodic perturbation to generate a topological phase in such a system out of a trivial (non-topological) phase. In this direction, we discuss the Floquet generation of topological modes in the BHZ model
(\ie realization of FFOTI) while starting from a non-topological phase~\cite{lindner11floquet}. The model Hamiltonian for the BHZ model reads~\cite{bernevig2006quantum}
\begin{align}
	H_{\rm BHZ} (\vect{k}) =& A \sin k_x  \ \sigma_x s_z +  A \sin k_y  \ \sigma_y s_0 \non\\
	&+ \left( M -4B +  2B \cos k_x +2B \cos k_y \right)  \sigma_z s_0 \ ,  \quad 
	\label{BHZ_model}
\end{align}
where, $A$, $B$, and $M$ represent the amplitude of the SOC, hopping, and crystal field splitting, respectively. The Pauli matrices $\vect{\sigma}$ and $\vect{s}$ act on the orbital and spin degrees of freedom, respectively. The static system becomes topological if $0 < M < 8B$. Afterward, we introduce the driving protocol as harmonic time-dependence in the onsite term as
\begin{equation}
	V(t)= V \cos \omega t \ \sigma_z s_0 \ .
\end{equation}
Here, $V$ and $\omega$ are the amplitude and frequency of the drive, respectively. Then the time-dependent system can either be studied employing the time-domain picture: by formulating the time evolution operator $U(t,0)$ [Eq.~(\ref{FloquetOptime})] or in the frequency-domain: by constructing the Floquet Hamiltonian $H_{\rm F}^\infty$ [Eq.~(\ref{FloquetHam})]. However, here we prefer the former scenario and compute the Floquet operator $U(T,0)$ to extract the features of the driven system.

\begin{figure}[]
	\centering
	\includegraphics[width=0.45\textwidth]{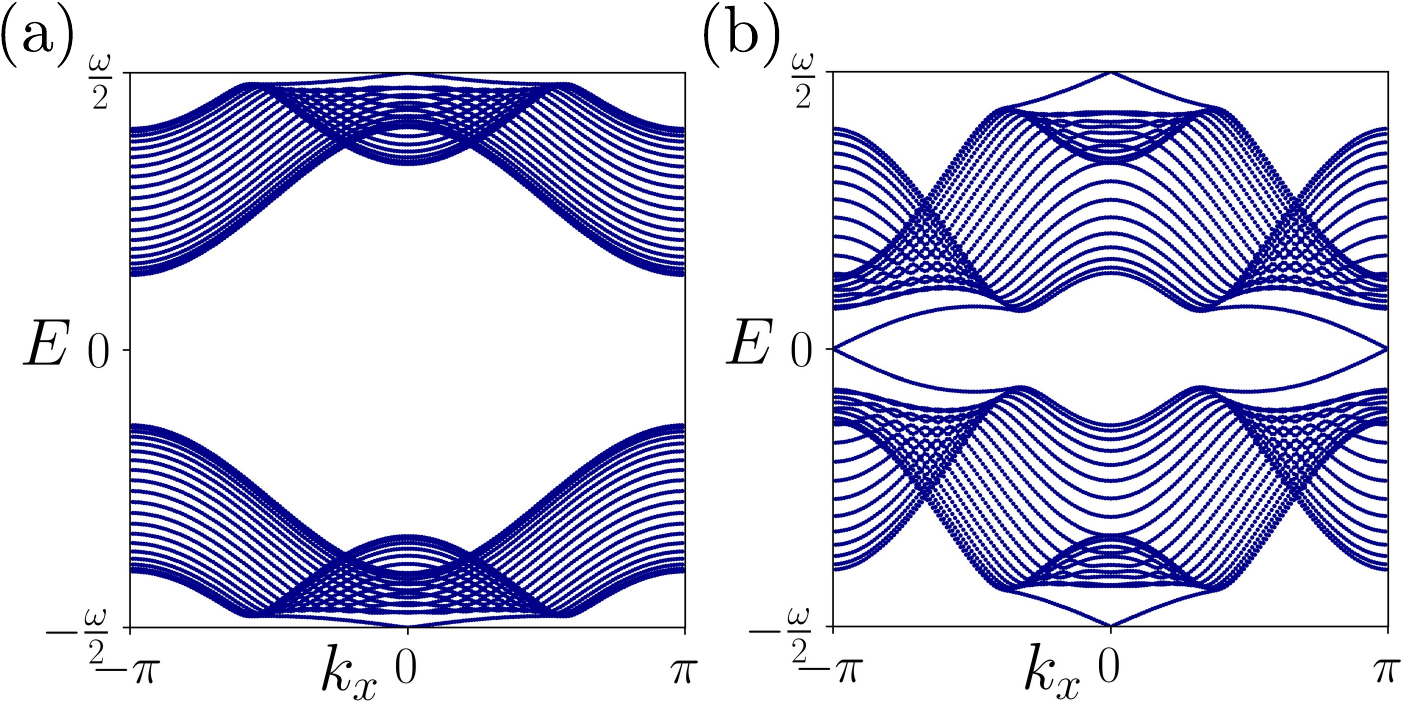}
	\caption{(a) We depict the quasi-energy spectra as a function of $k_x$ for a harmonically driven BHZ model, starting from the static non-topological regime with $M=-1.0$. The driving frequency for this case is chosen as $\Omega=3.0$. In panel (b), we repeat the same while starting from the topological regime with $M=1.0$. We choose $\omega=1.5$ for this case. The rest of the model parameters take the values: $A=B=0.2$ and $V=1.0$.
	}
	\label{BHZDriven}
\end{figure}

Case I: First, we consider the static undriven system to reside in the non-topological regime such that $M<0$. We consider ribbon geometry (\ie the system is finite along $y$-direction and infinite along $x$-direction) and diagonalize the Floquet operator. We depict the eigenvalue spectra $E$ in Fig.~\ref{BHZDriven} (a). This has been computed from the Floquet operator $U(T,0)$. One can identify a gapless dispersive mode at the quasi-energy $\pm\omega/2$ or $\pi$-gap. These $\pi$-modes are genuinely dynamical and possess no static analog. Also, such Floquet edge modes are chiral in nature.

Case II: On the other hand, if we start from the topological regime ($0<M<8B$), one can generate both $0$- and $\pi$-mode as depicted in Fig.~\ref{BHZDriven} (b). There can be other cases as well where multiple number of $0$- and  $\pi$-modes can appear simultaneously as well as  separately. Generation of similar $0$- and $\pi$-Floquet modes can be shown to exist in gapless systems \eg graphene~\cite{MikamiBW2016,Eckardt2015,Piskunow2014,Usaj2014}.


\section{Floquet generation of HOTIs} \label{Sec:FloquetHOTI} 
In this section, we discuss different driving protocols to generate the Floquet HOTI~(FHOTI) phase hosting higher-order boundary modes while starting from a trivial or a first-order topological phase.

\subsection{Perturbation kick in two dimensions} \label{Sec:Perturbation2D}

\begin{figure}[]
	\centering
	\includegraphics[width=0.3\textwidth]{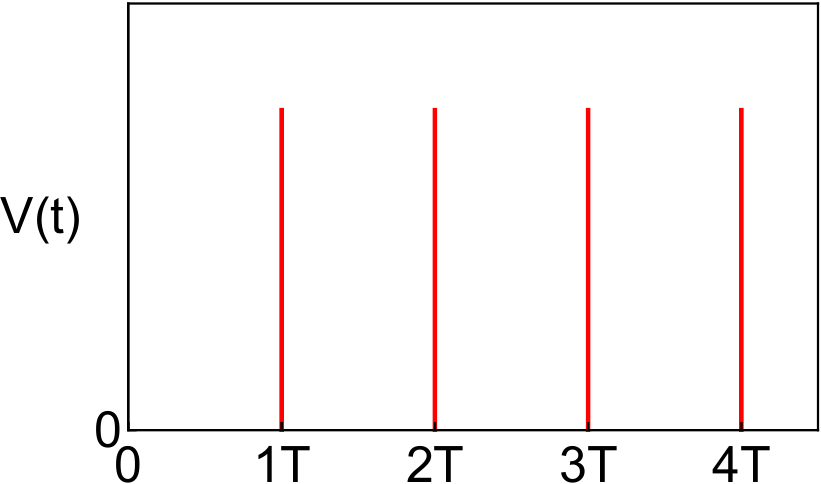}
	\caption{(a) A schematic portrayal of $C_4$ and $\mathcal{T}$ symmetry breaking periodic kick $V(t)$ [see Eq.~(\ref{Eq:periodickick})]. This type of drive protocol gives rise to 0D corner-localized modes at time $t=T$ [see Fig.~\ref{2DPertrbationKick2} (b)]. This figure is adapted from Ref.~\cite{Nag19} \copyright APS.
	}
	\label{2DPertrbationKick1}
\end{figure}

The Floquet SOTI~(FSOTI) hosting 0D corner modes can be generated employing a periodic kick in $C_4$ and $\mathcal{T}$ breaking Wilson-Dirac mass term while starting from a static QSHI~\cite{Nag19}. To this end, we begin with the Hamiltonian of 2D QSHI as
\begin{equation}~\label{Eq:QSHI_Hamiltonian}
	H_{\rm QSHI}= t_1 \sum^{2}_{j=1} \Gamma_j \sin k_j - t_0   \big[m- \sum^2_{j=1} \cos k_j \big] \Gamma_3
	\equiv {\bf N} ({\bf k}) \cdot {\boldsymbol \Gamma},
\end{equation}
where $k_j$ represents momentum along the $j^{\rm th}$ direction, $t_0$, $t_1$, and $m$ indicate the hopping amplitude, SOC strength, and the crystal field splitting, respectively. The $4 \times 4$ anti-commuting $\vect{\Gamma}$ matrices are given as: $\Gamma_1=\sigma_x s_z $, $\Gamma_2=\sigma_y s_0$, and $\Gamma_3=\sigma_z s_0$. The Pauli matrices $\vect{\sigma}$ and $\vect{s}$ operate on the orbital and spin degrees of freedom, respectively. The system represents a QSHI hosting propagating 1D edge modes when $0 < |m|< 2$~\cite{bernevig2006quantum}. We introduce the time-dependent perturbation in terms of a periodic kick in $C_4$ and $\mathcal{T}$ braking Wilson-Dirac~(WD) mass term as [see Fig.~\ref{2DPertrbationKick1} (a) for a schematic]
\begin{align}~\label{Eq:periodickick}
	V(t) = V_{xy} \Gamma_4 \sum^{\infty}_{r=1} \; \delta \left( t- r \; T \right),
\end{align}
where $V_{xy}=\Delta \left( \cos k_x - \cos k_y \right)$, $r$ is an integer, and $\Gamma_4=\sigma_x s_x$. We recover a static SOTI hosting 0D corner modes, if we consider the Hamiltonian $H_{\rm Stat}=H_{\rm QSHI}+ \Gamma_4 V_{xy}$. As $\{ H_{\rm QSHI} , \Gamma_4 \}=0$. The term $V_{xy} \Gamma_4$ breaks the $C_4=  e^{-i \frac{\pi}{4} \sigma_z s_z}$ as well as TRS $\mathcal{T}= i \sigma_0 s_y\mathcal{K}$ (with $K$ being the complex conjugation operation). However, it respects the combined $C_4 \mathcal{T}$ symmetry. One can show that this WD mass term induces a finite mass term proportional to $\Delta$ in the edge states of QSHI and then one can employ the Jackiw-Rebbi theorem to understand the appearance of the corner-localized zero-modes~\cite{jackiw1976solitons,ghosh2020floquet,ghosh2020floquet2}. 

\begin{figure}[]
	\centering
	\includegraphics[width=0.48\textwidth]{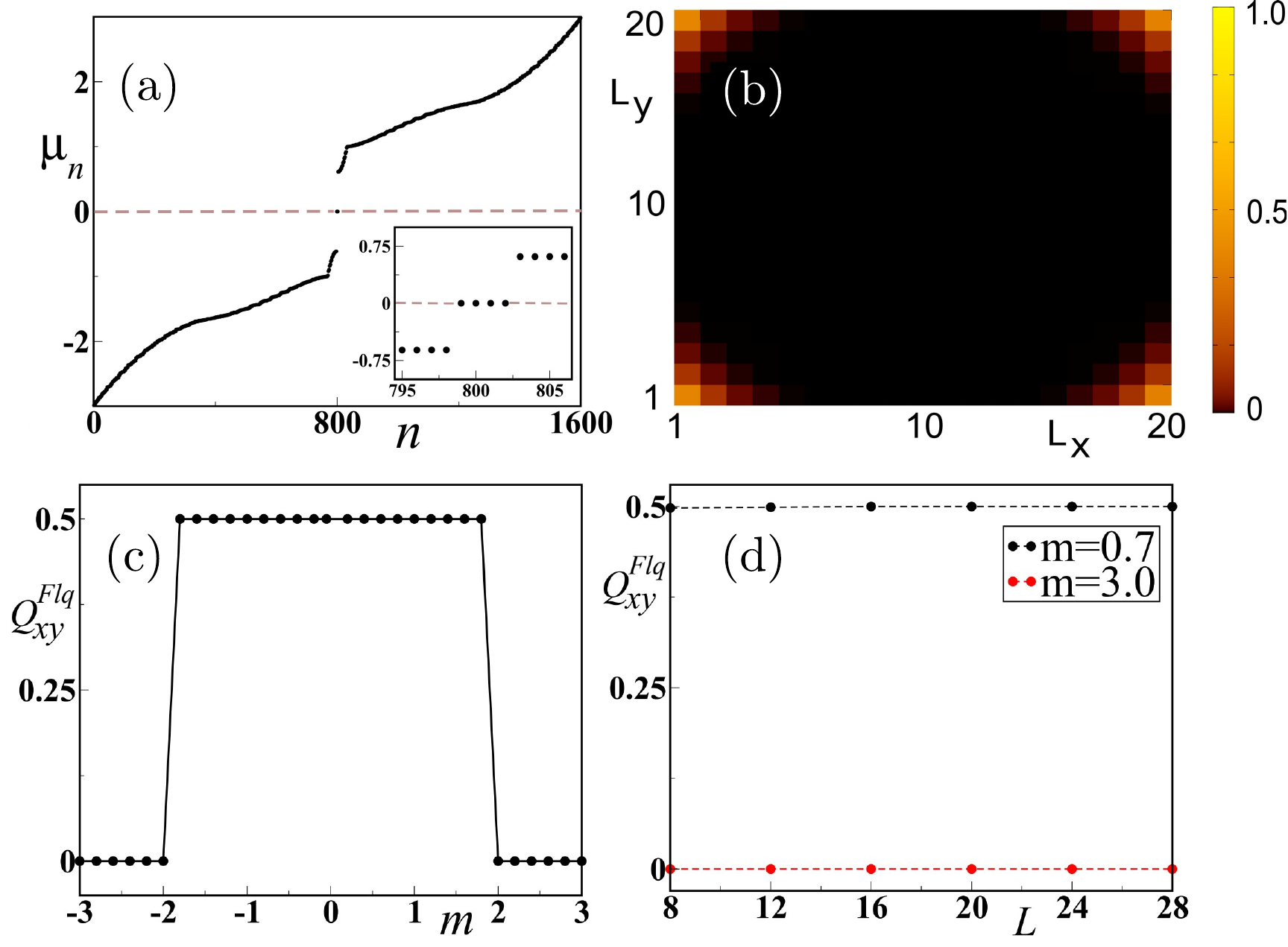}
	\caption{Engineering of a 2D FSOTI, hosting four zero-quasienergy corner modes ($\mu_n=0$). In panel (a), we depict the quasi-energy $\mu_n$ as a function of state index $n$. In the inset, we demonstrate the zoomed-in quasi-energy around $\mu_n=0$. We illustrate the LDOS distribution corresponding to the $\mu_n=0$ states of the Floquet operator $U(T)$ [see Eq.~(\ref{Eq:Floquetoperatorkick2D})] in panel (b). The model parameters are chosen as $t_0=t_1=m=1$, $\Delta=0.3$, and $T=0.628$. The corresponding frequency $\omega=2 \pi /T \approx 10 \gg t_{0,1}$. (c) Floquet quadrupolar moment ($Q^{\rm Flq}_{xy}$) is depicted as a function of $m$. The system consists of $20 \times 20$ lattice sites with the choice of origin at $(x_0,y_0)=(0,0)$. The system exhibits a FSOTI (trivial phase) for $|m|<2~(|m|>2)$ with $Q^{\rm Flq}_{xy}=0.5~(0.0)$. (d) The system size ($L$) independence of $Q^{\rm Flq}_{xy}$ is shown inside FSOTI ($m=1.0$) and trivial ($m=3.0$) phase. This figure has been adapted from Ref.~\cite{Nag19} \copyright APS.
	}
	\label{2DPertrbationKick2}
\end{figure}

To realize the FSOTI in the presence of the periodic kick [see Fig.~\ref{2DPertrbationKick1} (a) and Eq.~(\ref{Eq:periodickick})], we construct the Floquet operator in terms of the time-ordered (${\rm TO}$) product as
\begin{align}~\label{Eq:Floquetoperatorkick2D}
	U(T) &= {\rm TO} \exp \left[ -i \int_0^T \left[H_{\rm QSHI} + V(t) \right] dt \right]  \nonumber \\
	&= \exp(-i H_{\rm QSHI} \; T) \; \exp(-i V_{xy} \Gamma_4),
\end{align}
One can obtain the effective Floquet Hamiltonian ($H_{\rm Flq}$): $U(T)=\exp(-i H_{\rm Flq} T) \approx 1- i H_{\rm Flq} T + {\mathcal O} (T^2)$. In the high-frequency limit ($T \to 0$), we only keep the terms upto first-order in $T$ and obtain the effective Floquet Hamiltonian as
\begin{equation}~\label{Eq:FloquetHamiltonian}
	H_{\rm Flq} = \sum^3_{j=1} N_j ({\bf k}) \Gamma_j + V_{xy} \; \sum^3_{j=1} N_{j} ({\bf k}) \Gamma_{j4} + \frac{V_{xy}}{T} \; \Gamma_4 \ ,
\end{equation}
where $\Gamma_{jk}=[\Gamma_j, \Gamma_k]/(2i)$. Here, we further assume that $T, \Delta \to 0$, but $\Delta/T$ is finite. To numerically demonstrate that the driven system represents an FSOTI, we diagonalize the Floquet operator [Eq.~(\ref{Eq:Floquetoperatorkick2D})] employing OBC in both the $x$- and $y$-directions and obtain an eigenvalue equation of the form: $U(T) \; \ket{\phi_n}=\exp(-i \mu_n T) \; \ket{\phi_n}$; where $\ket{\phi_n}$ and $\mu_n$ are the Floquet state and the quasi-energy, respectively. We illustrate the quasi-energy spectra in Fig.~\ref{2DPertrbationKick2} (a) as a function of the state index $n$. The presence of the zero quasi-energy states is evident from the inset of Fig.~\ref{2DPertrbationKick2} (a). While we depict the LDOS corresponding to these zero quasi-enegy 
states in Fig.~\ref{2DPertrbationKick2} (b) and it is evident that these states are localized at the corners of the system. However, no $\pi$-modes appear for such drive. 

One can show that the 2D static SOTI exhibits quantized quadrupolar moment $Q_{xy}=0.5~({\rm mod}~1)$~\cite{benalcazar2017,wheeler2018many,kang2018many} [see Eq. (\ref{macroquadint}) for the definition]. To compute the Floquet quadrupolar moment $Q^{\rm Flq}_{xy}$, we first construct the matrix $U$ by columnwise arranging the eigenvectors $| \phi_n \rangle$ according to their quasi-energy 
$\mu_n$, such that $-\frac{\omega}{2} < \mu_n < 0$ while noticing that the Floquet operator has the quasi-energy window $\left(-\frac{\omega}{2},\frac{\omega}{2} \right)$. Afterward we follow the similar procedure as mentioned in Sec.~\ref{Subsection:BBHmodel} [see Eqs.~(\ref{macroquadint})-(\ref{Quadfinalint})] to compute $Q^{\rm Flq}_{xy}$. The quadrupolar moment $Q^{\rm Flq}_{xy}$ is depicted for this driven system as a function of $m$ in Fig.~\ref{2DPertrbationKick2} (c). One can observe that the FSOTI phase is obtained for $|m|<2$, exhibiting $Q^{\rm Flq}_{xy}=0.5$ while $Q^{\rm Flq}_{xy}=0$ represents a trivial system. One can further verify that the quantization of the Floquet quadrupolar moment does not depend upon the system size as illustrated in Fig.~\ref{2DPertrbationKick2} (d).

\vspace {-0.4cm}
\subsection{Step drive in two dimensions} \label{Sec:StepDrive2D}

One can also employ a periodic two-step drive protocol to generate the FSOTI in two dimensions~\cite{ghosh2021systematic}. This protocol allows us to realize the dynamical $\pi$-modes along with the concurrent $0$-modes. The driving protocol reads
\begin{align}
	H_{d \rm D}&= J_1' h_{1, 2 \rm D}(\vect{k}) \  ,   \quad \quad t \in \Bigg[0,\frac{T}{2}\Bigg] \ , \nonumber \\
	&=J_2' h_{2, 2 \rm D}(\vect{k}) \  , \quad \quad t \in \Bigg(\frac{T}{2},T \Bigg] \ ,
	\label{stepdrive2D}
\end{align}
where $J_1'$ and $J_2'$ carry the dimensions of energy. We set $\hbar=c=1$ and define two dimensionless parameters such as $(J_1,J_2)=(J_1' T, J_2' T)$; where $T$ represents the period of the drive and the corresponding driving frequency is given as $\Omega=2 \pi / T$. At the $i^{\rm th}$ step, the Hamiltonian of the system is represented by $J'_i h_{i, 2 \rm D}(\vect{k})$. We consider $h_{1, 2 \rm D}(\vect{k})=\sigma_z $ and $h_{2, 2 \rm D}(\vect{k})=\left( \cos k_x + \cos k_y \right) \sigma_z+\sin k_x \sigma_x s_z + \sin k_y \sigma_y + \alpha \left( \cos k_x - \cos k_y \right)  \sigma_x s_x $ to generate FSOTI hosting localized corner modes. The two Pauli matrices $\vect{\sigma}$ and $\vect{s}$ operate on orbital $(a,b)$ and spin $(\uparrow,\downarrow)$ degrees of freedom, respectively. The first Hamiltonian $J'_1 h_{1, 2 \rm D}(\vect{k})$ comprises of the on-site term only and respects all the necessary symmetries. On the other hand, the Hamiltonian in the second step $J'_2 h_{2, 2 \rm D}(\vect{k})$, incorporates all the hopping terms. When the term proportional to $\alpha$ is present (absent) in $J'_2 h_{2, 2 \rm D}(\vect{k})$, it breaks (respects) the TRS ($\mathcal{T}=i s_y \mathcal{K}$), the four-fold rotation ($C_4$) symmetry, and the mirror symmetries. The combined $C_4 \mathcal{T}$ symmetry is still preserved nonetheless.

Before moving to the dynamic case, we first discuss the static counterpart of this model. One may consider the following static Hamiltonian
\begin{align}
	H_{2 \rm D}^{\rm Static}(\vect{k})=& J_1' h_{1, 2 \rm D}(\vect{k}) + J_2' h_{2, 2 \rm D}(\vect{k}) \ , \label{ch5_2DStatic} 
\end{align}
where $H_{2 \rm D}^{\rm Static}(\vect{k})$ represents the Hamiltonian of a 2D QSHI hosting 1D helical gapless edge states, when $\alpha=0$ and $0< \lvert J_1' \rvert < 2 \lvert J_2' \rvert $~\cite{bernevig2006quantum,konig2007quantum}. However, for $\alpha \neq 0$, the edge states of the QSHI attain a mass term proportional to $\alpha$ and become massive in such a way that two intersecting edges carry opposite mass terms. One can use a similar line of arguments as discussed in Sec.~\ref{Sec:Perturbation2D} to understand that the Hamiltonian $H_{2 \rm D}^{\rm Static}(\vect{k})$ [Eq.~(\ref{ch5_2DStatic})] represents a 2D SOTI hosting localized 0D corner modes.

\begin{figure}[]
	\centering
	\includegraphics[width=0.48\textwidth]{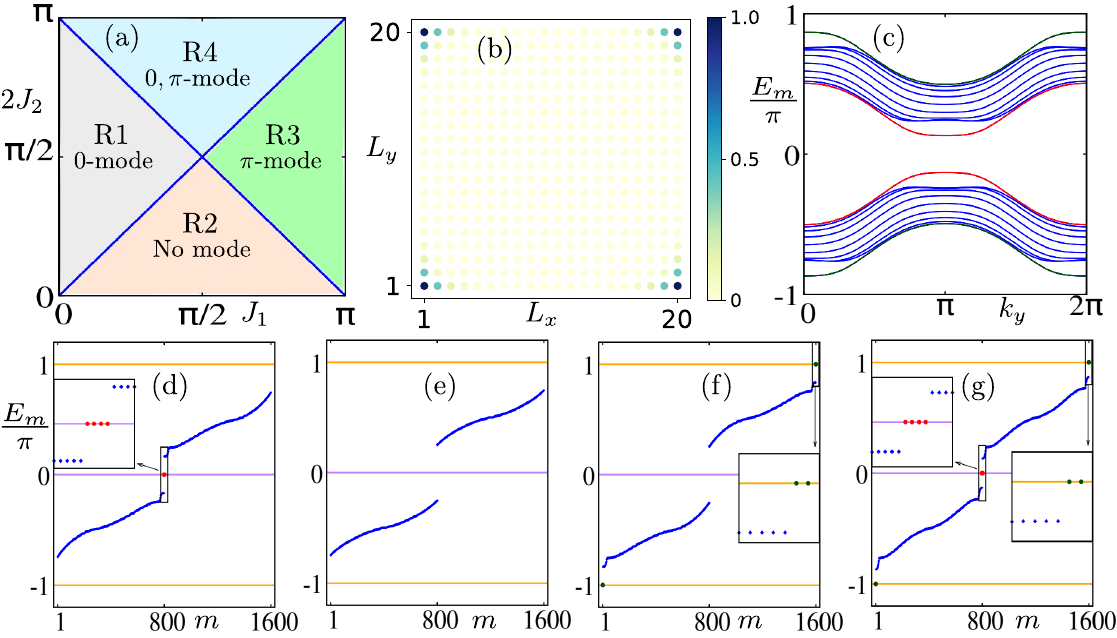}
	\caption{Floquet generation of 2D SOTI using two-step drive. We depict the phase diagram in the $J_1 \mhyphen J_2$ plane in panel (a). The blue lines separate four phases: R1, R2, R3, and R4. In panel (b), the LDOS for the quasi-energies $E_m=0,~\pm \pi$ is demonstrated as a function of the system dimensions. In this case, the system resides in R4. (c) Quasi-energy spectrum $E_m$ is illustrated for a system obeying PBC in one direction, as a function of $k_y$ for R4. The edge modes are gapped out at both quasi-energy $0$ and $\pi$. In panels (d), (e), (f), and (g), we show the quasi-energy spectra $E_m$ as a function of the state index $m$ for the R1, R2, R3, and R4 phases, respectively. The model parameters are chosen as $(J_1,2 J_2)=\left[ \left(\frac{\pi}{4},\frac{\pi}{2}\right),~\left(\frac{\pi}{2},\frac{\pi}{4}\right),~\left(\frac{3\pi}{4},\frac{\pi}{2}\right), ~\left(\frac{\pi}{2},\frac{3 \pi}{4}\right) \right] $ for R1, R2, R3, and R4, respectively. This figure is adapted from Ref.~\cite{ghosh2021systematic} \copyright APS.
	}
	\label{2Dstep}
\end{figure}

Following the step drive protocol, we obtain the Floquet operator $U_{2 \rm D}(\vect{k},T)$ as
\begin{align}\label{ch5_U2Dstep}
	U_{2 \rm D}(\vect{k},T)= \exp\left(-i \frac{J_2}{2} h_{2, 2 \rm D}(\vect{k}) \right) \exp\left(-i \frac{J_1}{2} h_{1, 2 \rm D}(\vect{k})  \right)\ , \qquad
\end{align}
where one can express $U_{2 \rm D}(\vect{k},T)$ as $U_{2 \rm D}(\vect{k},T)=f_{2 \rm D}(\vect{k})  \mathbb{I}+i g_{2 \rm D}(\vect{k})$, such that 
\begin{widetext}
	\begin{align}
		f_{2 \rm D}(\vect{k})=& \cos\left(\gamma_{2 \rm D}(\vect{k})  \frac{J_1}{2}\right) \cos\left( \lambda_{2 \rm D}(\vect{k}) \frac{J_2}{2} \right) - \sin\left(\gamma_{2 \rm D}(\vect{k}) \frac{J_1}{2}\right) \sin\left( \lambda_{2 \rm D}(\vect{k}) \frac{J_2}{2} \right)  \chi_{2 \rm D}(\vect{k}) \ , \label{ch5_fk2Dstep}  \\
		g_{2 \rm D}(\vect{k})=& - \frac{1}{\gamma_{2 \rm D}(\vect{k}) \lambda_{2 \rm D}(\vect{k})}  \sin\left(\gamma_{2 \rm D}(\vect{k})  \frac{J_1}{2}\right) \sin\left( \lambda_{2 \rm D}(\vect{k})  \frac{J_2}{2} \right) \eta_{2 \rm D}(\vect{k}) -  \sin\left(\gamma_{2 \rm D}(\vect{k})  \frac{J_1}{2}\right) \cos \left( \lambda_{2 \rm D}(\vect{k})  \frac{J_2}{2} \right) \frac{h_{1, 2 \rm D}(\vect{k})}{\gamma_{2 \rm D}(\vect{k})} \non \\
		&-  \cos\left(\gamma_{2 \rm D}(\vect{k})  \frac{J_1}{2}\right) \sin \left( \lambda_{2 \rm D}(\vect{k})  \frac{J_2}{2} \right) \frac{h_{2, 2 \rm D}(\vect{k})}{\lambda_{2 \rm D}(\vect{k})} \  \label{ch5_gk2Dstep} \ .
	\end{align}
\end{widetext}
The implicit $T$ dependence in $f_{2 \rm D}(\vect{k})$ and $g_{2 \rm D}(\vect{k})$ are suppressed for brevity. We define $\gamma_{2 \rm D}(\vect{k})=\lvert h_{1, 2 \rm D}(\vect{k}) \rvert$, $\lambda_{2 \rm D}(\vect{k})=\lvert h_{2, 2 \rm D}(\vect{k}) \rvert$, $\chi_{2 \rm D}(\vect{k})=\frac{\cos k_x +\cos k_y}{\gamma_{2 \rm D}(\vect{k}) \lambda_{2 \rm D}(\vect{k})}$, and $\eta_{2 \rm D}(\vect{k})= \sin k_x \sigma_y s_z - \sin k_y \sigma_x + \alpha \left( \cos k_x - \cos k_y \right) \sigma_y s_x $. The eigenvalue equation for $U_{2 \rm D}(\vect{k},T)$: $U_{2 \rm D}(\vect{k},T) \ket{\Psi} = e^{i E(\vect{k})} \ket{\Psi}$, provides us with the following condition
\begin{equation}
	\cos E(\vect{k})=f_{2 \rm D}(\vect{k}) \ ,
\end{equation}
while the $E(\vect{k})$'s are two-fold degenerate. At $\vect{k}=\vect{k}^*=(0,0)~{\rm or}~(\pi,\pi)$, the band gap closes when $f_{2 \rm D}(\vect{k}^*)=\pm 1$. Note that, at $\vect{k}=\vect{k}^*=(0,0)~{\rm or}~(\pi,\pi)$, one finds that $\gamma_{2 \rm D}(0,0)=\gamma_{2 \rm D}(\pi,\pi)=1$, $\lambda_{2 \rm D}(0,0)=\lambda_{2 \rm D}(\pi,\pi)=2$, and $\chi_{2 \rm D}(0,0)=-\chi_{2 \rm D}(\pi,\pi)=1$. Thus, one may express $f_{2 \rm D}(\vect{k})$ in terms of a single cosine function such that $f_{2 \rm D}(\vect{k}^*)=\cos \left[ \gamma_{2 \rm D}(\vect{k}^*)  \frac{J_1}{2} \pm \lambda_{2 \rm D}(\vect{k}^*) \frac{J_2}{2} \right]$. The gap closing at these special momenta plays a pivotal role in finding the topological phase boundaries, and at these points, one can write Eq. (\ref{ch5_fk2Dstep}) in a compact form as 
\begin{equation}\label{ch5_cosphase2D}
	\cos \left( \frac{J_1}{2} \pm J_2  \right) = \cos n \pi \ ,
\end{equation}
where $n \in \mathbb{Z}$. Using Eq.~(\ref{ch5_cosphase2D}), one can obtain the gap closing conditions in terms of $J_1$ and $J_2$ as
\begin{equation}\label{ch5_phasestep2D}
	\lvert J_2 \rvert  = \frac{\lvert J_1 \rvert}{2} + n \pi \ .
\end{equation}
Here, Eq.~(\ref{ch5_phasestep2D}) provides us with the topological phase boundaries between various Floquet phases, as shown in Fig.~\ref{2Dstep} (a). The phase diagram encapsulates four fragments- region 1~(R1) hosting only $0$-mode, region 2~(R2) without any modes \ie trivial, region 3~(R3) hosting only $\pi$-mode, and region 4~(R4) accommodating both $0$- and $\pi$-modes. To examine the presence of the corner modes numerically, we diagonalize the Floquet operator [Eq.~(\ref{ch5_U2Dstep})] employing OBC in both $x$- and $y$-directions and illustrate the LDOS for the quasi-states with quasi-energy $E_m=0,~\pm \pi$ in Fig.~\ref{2Dstep} (b). We depict the quasi-energy spectra for the phases R1, R2, R3, and R4 in Figs.~\ref{2Dstep} (d), (e), (f), and (g), respectively. Here, we would like to mention that, instead of considering a two-step drive protocol [Eq.~(\ref{stepdrive2D})], one may also employ a three-step drive protocol as described in Refs.~\cite{Huang2020,GhoshDynamical2022}. However, the three-step drive protocol would also produce similar features as obtained for two-step drive due to the fact that the form of $f_{2 \rm D}(\vect{k}^*)$ remains unchanged in both the cases.

\subsection{Mass kick in two dimensions}

\begin{figure}[]
	\centering
	\includegraphics[width=0.45\textwidth]{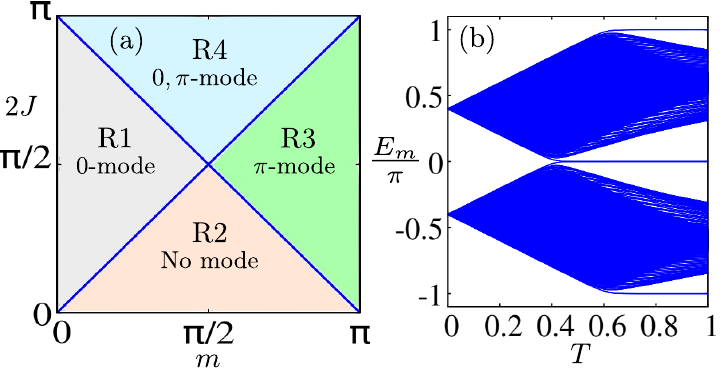}
	\caption{(a) We depict the phase diagram in the $m \mhyphen J$ plane for the mass kick drive protocol. The phase diagram is divided into four parts: R1, R2, R3, and R4. In panel (b), we illustrate the quasi-energy spectra $E_m$ as a function of time period $T$. The appearance of $0$- and $\pi$-modes can be identified with varying $T$. The model parameters are chosen as $(m,J')=(0.4 \pi, 0.5 \pi)$. This figure is adapted from Ref.~\cite{ghosh2021systematic} \copyright APS.
	}
	\label{2Dmasskick}
\end{figure}

Having demonstrated the step drive protocol to generate the 2D FSOTI phase hosting both $0$- and $\pi$-modes, we discuss another protocol namely the periodic kick protocol or the mass kick protocol~\cite{ghosh2021systematic} to generate the same. Between two successive kicks, we employ the Hamiltonian $\mathcal{H}_{2 \rm D}=J^{\prime} h_{2, 2 \rm D}(\vect{k})$ [see Eq.~(\ref{stepdrive2D})], with $J'$ carrying the dimension of energy. Similar to the step drive case, here also we employ two dimensionless parameters $J=J'T$ and $m$ to control the drive. Then, we introduce the mass kick protocol as
\begin{equation} 
	m_0(t)=m \ h_{1, 2 \rm D} \sum_{r=1}^{\infty} \delta(t-rT) \ , 
	\label{kick1sys2D}
\end{equation}
where, $m$, $t$, and $T$ denote the kicking parameter's strength, time, and the time-period of the drive, respectively. The Floquet operator, $U_{2 \rm D}(\vect{k},T)$ for this drive reads as
\begin{align}
	U_{2 \rm D}(\vect{k},T)&={\rm TO} \exp \left[-i\int_{0}^{T}dt\left(\mathcal{H}_{2 \rm D}({\vect{k}})+m_0(t) \right)\right] \ ,  \nonumber \\
	&= \exp(-i \mathcal{H}_{2 \rm D}({\vect{k}}) T)~\exp(-i m~h_{1, 2 \rm D})\ .
	\label{fomasskick2D}
\end{align}
One obtains the static counterpart of this mass kick protocol by considering a Hamiltonian of the form $H_{2 \rm D}^{\rm Static}(\vect{k})= m \ h_{1, 2 \rm D} + J' h_{2, 2 \rm D}(\vect{k})$. Note that, the step drive protocol [Eq. (\ref{stepdrive2D})] mimics the mass kick protocol [Eq. (\ref{kick1sys2D})] if one considers an infinitesimal duration for the first step Hamiltonian $J'_1 h_{1, 2 \rm D}$ of the step drive protocol. Following the similar line of arguments as discussed earlier for the step drive, the gap-closing conditions can be obtained for the mass kick protocol as
\begin{align}
	2 \lvert J \rvert =& \lvert m \rvert  + n \pi  \label{phasekick2D}\ ,
\end{align}
The phase diagram for the mass kick resembles that of the step-drive and we illustrate the same in Fig.~\ref{2Dmasskick} (a). The eigenvalue spectra also exhibit a qualitatively similar nature to that of the step drive protocol as shown in Figs.~\ref{2Dstep} (d), (e), (f), and (g) for the phases R1, R2, R3, and R4, respectively.

The difference between the topological phase boundary equations [Eqs. (\ref{ch5_phasestep2D}) and (\ref{phasekick2D})] of the step drive and periodic mass kick drive is the absence of $T$ in the latter case. Although $T$ is multiplied with both the driving strength in case of the step drive. However, in the case of periodic mass kick, the term $m$ in the right-hand side of Eq.~(\ref{phasekick2D}) is not coupled to $T$. This minute difference compared to the step drive allows us to investigate a frequency-driven topological phase transition for this case [see Fig.~\ref{2Dmasskick} (b)]. In particular, we choose $(m,J')=(0.4 \pi,0.5 \pi)$ so that the system lies in the phase R2 (without any modes) and increases (decreases) the time-period $T$ (frequency $\Omega$). First, we cross through R1, where we obtain only the $0$-modes, and then R4, where we observe the appearance of both the $0$- and $\pi$-modes. Thus, the periodic kick protocol can also mediate the frequency-driven topological phase transition~\cite{nag2021anomalous,ghosh2021systematic}.

\subsection{Laser irradiation in two dimensions}
\begin{figure}[]
    \centering
    \includegraphics[width=0.47\textwidth]{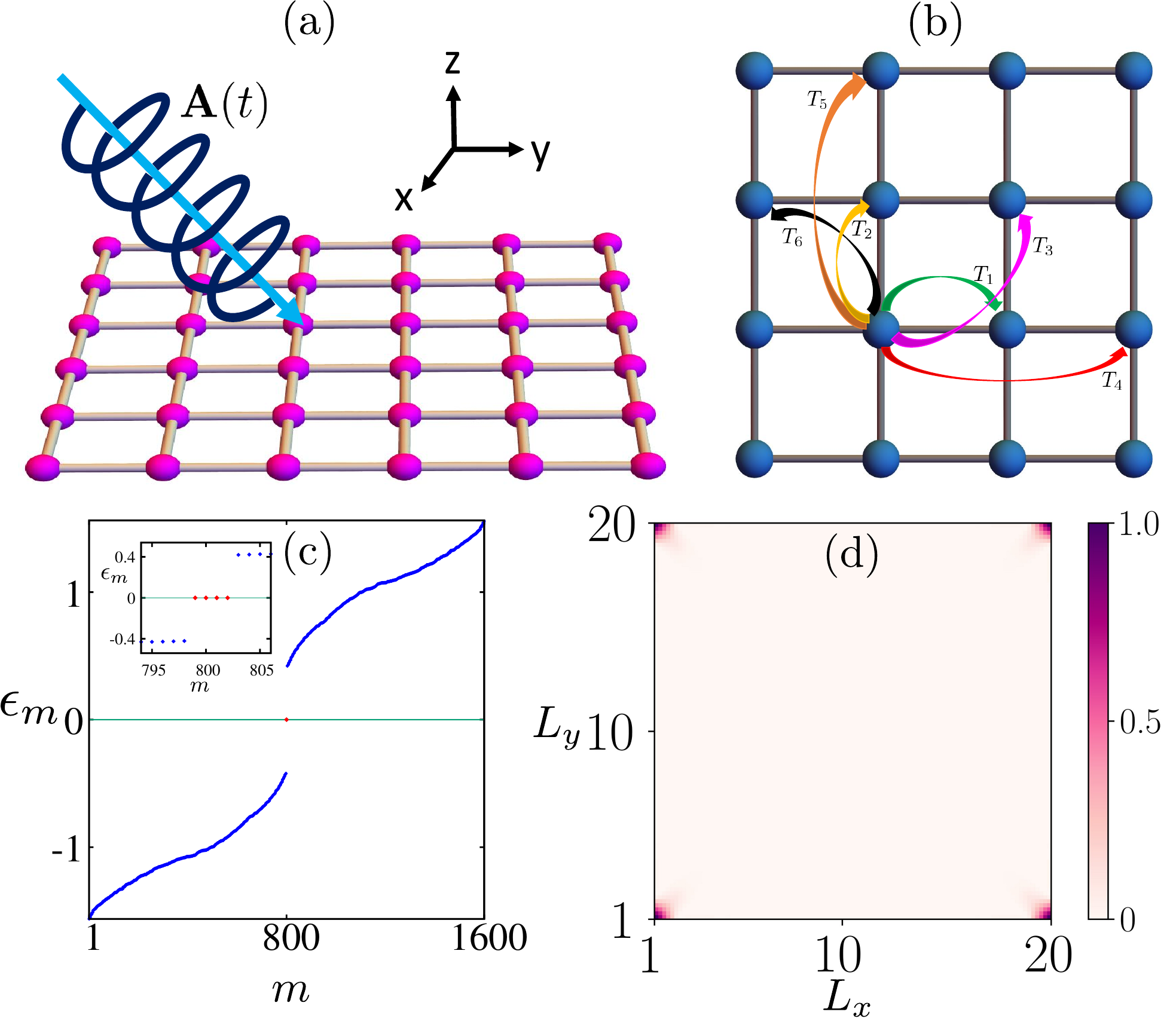}
    \caption{(a) We depict a schematic representation of our setup in presence of an external circularly polarized laser irradiation. In panel (b), we illustrate different drive-generated higher-order hoppings. Here, the nearest-neighbor renormalized hoppings are represented by $T_1$, and $T_2$, while the drive-induced high-order hopping parameters in different directions are denoted by $T_3$, $T_4$, $T_5$, and $T_6$. In panel (c), we demonstrate the quasi-energy spectrum as a function of the quasi-states index $m$. The four corner-state eigenvalues are denoted by red dots in the inset. (d) The LDOS distribution is depicted as a function of the system dimension choosing $E_m=0$. One can clearly identify that the zero-quasienergy states are populated at the four corners of the system. This figure is 
    adapted from Ref.~\cite{Ghosh2020} \copyright APS.
    }
    \label{2Dlaser1}
\end{figure}

We now employ an experimentally feasible driving protocol (laser irradiation) to study the effect of periodic drive on HOTIs~\cite{Ghosh2020}. To implement external irradiation, we consider a static semimetallic system based on a 2D square lattice. The Hamiltonian for such a system reads as
\begin{equation}\label{HSM}
	H_{\rm SM}=\sum_{j,k}^{}\left[c^\dagger_{j,k} T_x c_{j+1,k}+c^\dagger_{j,k} T_y c_{j,k+1} +\textrm{h.c.} \right]\ ,
\end{equation}
where
\begin{equation}
	T_x=\frac{i t_1}{2} \Gamma_1+\frac{ t_2}{2} \Gamma_3 \ , \quad T_y=\frac{i t_1}{2} \Gamma_2+\frac{ t_2}{2} \Gamma_3 \ , 	
\end{equation}
with $t_{1,2}$ are the amplitudes of nearest-neighbour hopping. We employ the following basis: $c_{j,k} = \{c_{A\ua},c_{B\ua},c_{A\da},c_{B\da}\}^{T}$. The $4 \times 4$ $\vect{\Gamma}$ matrices are given as $\Gamma_1=\sigma_3 \tau_1$, $\Gamma_2=\sigma_0 \tau_2$, and $\Gamma_3=\sigma_0 \tau_3$. The Pauli matrices ${\boldsymbol \tau}$ and ${\boldsymbol \sigma}$ act on the orbital $(A, B)$ and 
spin $(\uparrow,\downarrow)$ degrees of freedom, respectively. The HOTI phase is observed, if we add a WD mass term $H_B=\frac{\Delta}{2} \sum_{j,k}^{}\left[c^\dagger_{j,k} \sigma_1 \tau_1 c_{j+1,k} - c^\dagger_{j,k} \sigma_1 \tau_1 c_{j,k+1} +\textrm{h.c.} \right]$~\cite{schindler2018higher,Seshadri2019,agarwala2019higher,Nag19} with the semimetallic Hamiltonian $H_\textrm{SM}$ [Eq. (\ref{HSM})]. 
The total Hamiltonian $H=H_\textrm{SM}+H_B$ represents a static SOTI hosing in-gap localized corner modes~\cite{schindler2018higher,agarwala2019higher,Nag19}.

We depict the schematic setup of our system in the presence of circularly polarized laser irradiation in Fig.~\ref{2Dlaser1} (a). Laser irradiation can be obtained by considering the vector potential of the form: $\mathbf{A} (t)=\mathbf{A} (\cos(\omega t),\sin(\omega t))$; with $\omega$ being the driving frequency. To avoid any spatial dependency of the irradiation, we assume the beam spot of the external laser to be larger than that of the system. In comparison to the linearly polarized light, we choose circularly polarized light since the latter (former) breaks (preserves) TRS. Apparently, it appears that the breaking of TRS is important to achieve non-trivial phases in the driven system~\cite{arijitSilicene,Piskunow2014,Usaj2014}. We first consider a high-frequency limit such that $\omega \gg$ bandwidth of the system. Afterward, we employ the BW perturbation theory and obtain an effective Floquet Hamiltonian as a power series in $1/\omega$~\cite{MikamiBW2016}. However, we keep only the leading order term in the series as it contributes significantly to the emergent important physical phenomena. The BW effective Hamiltonian reads as~\cite{MikamiBW2016}
\begin{equation}
	\mathcal{H}_{\textrm{BW}}=\sum_{r=0}^{\infty} \mathcal{H}_{\textrm{BW}}^{(r)}\ ,
	\label{BWH}
\end{equation}
with
\begin{align}
	\mathcal{H}_{\textrm{BW}}^{(0)}&=\mathcal{H}_0, \quad \mathcal{H}_{\textrm{BW}}^{(1)} = \sum_{n \neq 0} \frac{\mathcal{H}_{-n}\mathcal{H}_n}{n\omega}, \quad \mathcal{H}_{\textrm{BW}}^{(2)} = \mathcal{O}\left(\frac{1}{\omega^2}\right)\ . 
\end{align}
The Fourier components $\mathcal{H}_n$'s are provided in Eq.~(\ref{FourierCompH}). Therefore, using the BW expansion [Eq. (\ref{BWH})], we compute the effective Floquet Hamiltonian for our driven system such that
\begin{widetext}
\vspace{-0.3cm}
	\begin{align}
		\mathcal{H}_{\textrm{BW}}^{(0)} &=\sum_{j,k}^{}\Big[c^\dagger_{j,k} T_1 c_{j+1,k}+c^\dagger_{j,k} T_2 c_{j,k+1} +\textrm{h.c.} \Big]\ , \non\\    
		\mathcal{H}_{\textrm{BW}}^{(1)}&=\sum_{j,k}^{}c^\dagger_{j,k} M c_{j,k}+\sum_{j,k}^{}\Big[c^\dagger_{j,k} T_3 c_{j+1,k+1} +c^\dagger_{j,k} T_4 c_{j+2,k} +c^\dagger_{j,k} T_5 c_{j,k+2} +c^\dagger_{j,k} T_6 c_{j-1,k+1} +\textrm{h.c.} \Big]\ , 
		\label{FHam}    
	\end{align}
	with
	\begin{align}
		M&=J_2\big(t_1^2+t_2^2+\Delta^2 \big) \sigma_0\tau_0\ , \non\\
		T_1&= \frac{\mathcal{J}_0(A)}{2}\big(i t_1 \sigma_3\tau_1  + t_2 \sigma_0 \tau_3+\Delta \sigma_1 \tau_1 \big)\ , \non \\
		T_2&= \frac{\mathcal{J}_0(A)}{2}\big(i t_1 \sigma_0\tau_2  + t_2 \sigma_0 \tau_3-\Delta \sigma_1 \tau_1\big)\ , \non \\
		T_3&=  J_{c1}\big(t_2^2-\Delta^2\big)\sigma_0\tau_0+J_{s1}\big(t_1^{2}\sigma_3\tau_3+it_1t_2\sigma_3\tau_2+it_1t_2\sigma_0\tau_1+2t_2\Delta\sigma_1\tau_2+it_1\Delta\sigma_2\tau_0-it_1\Delta\sigma_1\tau_3\big)\ ,  \non \\
		T_4&= J_1\big(t_2^2-t_1^2+\Delta^2 \big) \sigma_0\tau_0\ , \non\\
		T_5&= J_1\big(t_2^2-t_1^2+\Delta^2 \big)\sigma_0\tau_0\ ,     \non \\
		T_6&=  J_{c2}\big(t_2^2-\Delta^2\big)\sigma_0\tau_0-J_{s2}\big(t_1^{2}\sigma_3\tau_3+it_1t_2\sigma_3\tau_2-it_1t_2\sigma_0\tau_1-2t_2\Delta\sigma_1\tau_2+it_1\Delta\sigma_2\tau_0+it_1\Delta\sigma_1\tau_3\big)\ ,  
		\label{hotih}
	\end{align}
	where,
	\begin{align}
		J_1&=\sum_{n \neq0}\frac{(-1)^n\mathcal{J}_n^2(A)}{4n\omega},\quad  &J_{c1}&=\sum_{n \neq 0} \frac{(-1)^n \mathcal{J}_n^2(A) \cos\big(\frac{n \pi}{2}\big) }{2n\omega},\quad  &J_{s1}&=\sum_{n \neq 0} \frac{(-1)^n \mathcal{J}_n^2(A) \sin\big(\frac{n \pi}{2}\big) }{2n\omega}  \non\\
		J_2&=\sum_{n \neq 0}\frac{\mathcal{J}_n^2(A)}{n\omega}, &J_{c2}&=\sum_{n \neq 0} \frac{\mathcal{J}_n^2(A) \cos\big(\frac{n \pi}{2}\big) }{2n\omega}, &J_{s2}&=\sum_{n \neq 0} \frac{ \mathcal{J}_n^2(A)\sin\big(\frac{n \pi}{2}\big) }{2n\omega}. 
	\end{align}
	Here, $\mathcal{J}_n$ is the Bessel function of the first kind and $A$ is the amplitude of the drive.
\vspace{+0.2cm}
\end{widetext}
From Eq. (\ref{HSM}), one can observe that the $0^{\rm th}$-order Hamiltonian $\mathcal{H}_{0}$ is analogous to the unperturbed static Hamiltonian with renormalized hopping amplitudes $T_1$ and $T_2$. While the term with $\mathcal{O}(1/\omega)$ encompasses new drive-generated long-range hoppings ~\cite{MikamiBW2016,arijitSilicene,Usaj2014}. These terms are represented by $T_3$, $T_4$, $T_5$, and $T_6$. We depict these hopping parameters schematically in Fig.~\ref{2Dlaser1} (b).

We now discuss the primary numerical results obtained employing laser irradiation. The FSOTI phase is identified by the emergence of zero-quasienergy Floquet corner modes~\cite{Huang2020,Martin2019,Bomantara2019}. We illustrate the eigenvalue spectrum of the BW Hamiltonian in Fig.~\ref{2Dlaser1} (c). One can clearly identify the presence of four zero-quasienergy states (represented by the red dots) from the inset of Fig.~\ref{2Dlaser1} (c). The signatures of these zero-quasienergy Floquet corner modes can be understood via the LDOS computed at quasienergy $E=0$. We show the corrsponding LDOS associated with the Floquet corner modes in Fig.~\ref{2Dlaser1} (d) as a function of the two spatial dimensions $L_x$ and $L_y$. The zero-quasienergy states are localized at the four corners of the system. Thus, the 0D Floquet corner modes are robust against the high-frequency laser irradiation drive and pinned at zero quasienergy in a driven setup. We also topologically characterize these Flqouet corner modes by computing the quadrupole moment, $Q_{xy}$. For that, we follow a similar procedure as discussed in Sec.~\ref{Subsection:BBHmodel}. However, no driving strength $A$ dependent phase transition is observed rather $Q_{xy}$ always exhibits $0.5$ (mod 1) for any value of $A$. Nevertheless, the Floquet corner modes that appear in this driven system 
are different in nature when compared to the static ones. In a driven system, the manifestation of topological modes is due to the virtual photon transitions between different Floquet sub-bands~\cite{Eckardt2015}. 

\begin{figure}[]
    \centering
    \includegraphics[width=0.49\textwidth]{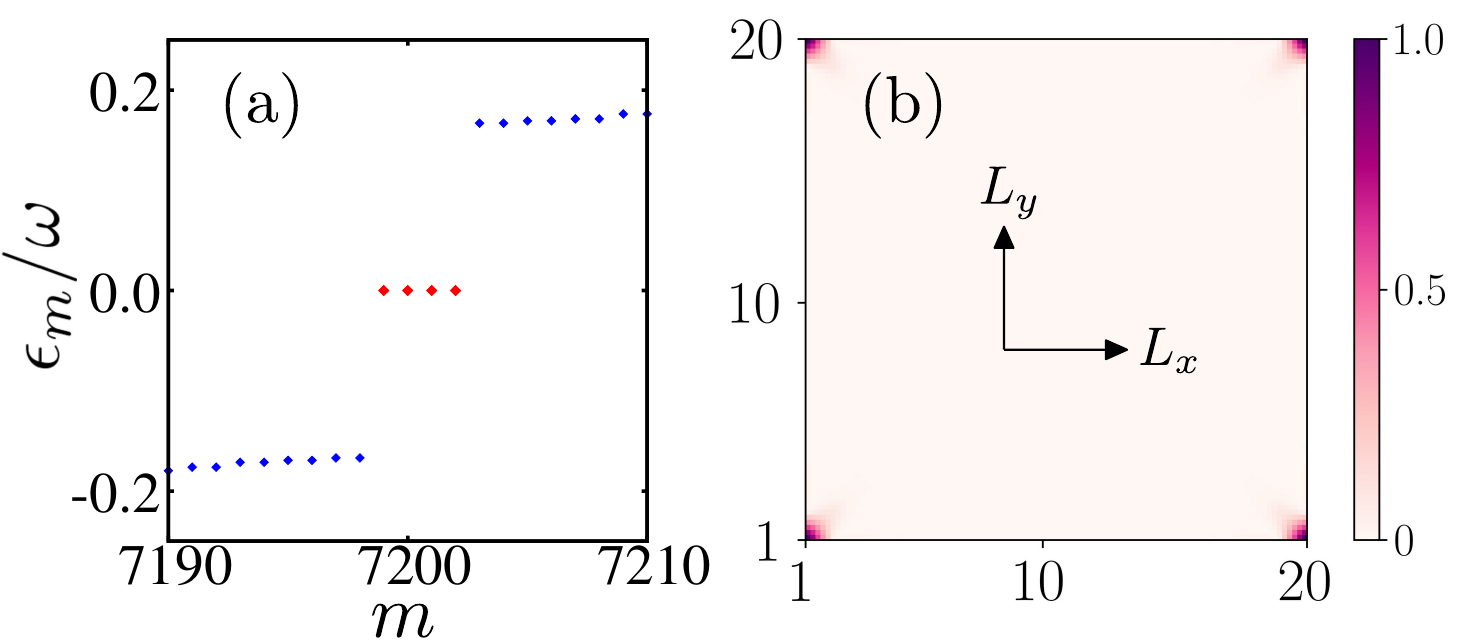}
    \caption{We depict the quasi-energy spectrum highlighting around quasi-energy $0$ (modulo $\omega$) in panel (a). In panel (b), the LDOS at $\epsilon_m=0$ (modulo $\omega$) for $A\neq 0$ 
    is shown. Here, we choose $A=0.5$ and $\omega=3.0$. This figure is adapted from Ref.~\cite{Ghosh2020} \copyright APS.
    }
    \label{2Dlaser2}
\end{figure}

Having demonstrated the emergence of the FHOTI phase in the high-frequency limit, we now investigate the system when the frequency of the driving photon field is comparable to the bandwidth of the system, such that $\omega \sim$ bandwidth. In such a limit, one can obtain the dressed corner modes. However, in that limit, any perturbation theory breaks down~\cite{Usaj2014,Piskunow2014,YangPRL2019}. Thus, one may consider a truncated Hamiltonian up to some Floquet zone sectors from the infinite-dimensional Floquet Hamiltonian $H_{\rm F}^{\infty}$ 
[Eq. (\ref{FloquetHam})]. In particular, we choose up to the $m=4$ zone, and the curtailed Hamiltonian reads as
\begin{widetext}
\begin{equation}\label{midHam}
    \tilde{H}_F=\begin{pmatrix}
        \mathcal{H}_0-3\omega & \mathcal{H}_1 & \mathcal{H}_2 & \mathcal{H}_3 & 0 & 0 & 0\\
        \mathcal{H}_{-1} & \mathcal{H}_0-2\omega & \mathcal{H}_1 & \mathcal{H}_2 &\mathcal{H}_3 & 0 & 0\\
        \mathcal{H}_{-2} & \mathcal{H}_{-1} & \mathcal{H}_0-\omega & \mathcal{H}_1 & \mathcal{H}_2 &\mathcal{H}_3 & 0 \\
        \mathcal{H}_{-3} & \mathcal{H}_{-2} & \mathcal{H}_{-1} & \mathcal{H}_0 & \mathcal{H}_1 & \mathcal{H}_2 &\mathcal{H}_3 \\
        0&\mathcal{H}_{-3} & \mathcal{H}_{-2} & \mathcal{H}_{-1} & \mathcal{H}_0+\omega & \mathcal{H}_1 & \mathcal{H}_2 \\
        0&0&\mathcal{H}_{-3} & \mathcal{H}_{-2} & \mathcal{H}_{-1} & \mathcal{H}_0+2\omega & \mathcal{H}_1 \\
        0&0&0&\mathcal{H}_{-3} & \mathcal{H}_{-2} & \mathcal{H}_{-1} & \mathcal{H}_0+3\omega 
    \end{pmatrix} \ ,
\end{equation}
where
 \begin{align}
	\mathcal{H}_0 &=\sum_{j,k}^{}\Big[c^\dagger_{j,k} \mathcal{T}_1 c_{j+1,k}+c^\dagger_{j,k} \mathcal{T}_2 c_{j,k+1} +\textrm{h.c.} \Big]\ , \quad \mathcal{H}_1 =\sum_{j,k}^{}\Big[c^\dagger_{j,k} \mathcal{T}_3 c_{j+1,k}+c^\dagger_{j,k} \mathcal{T}_4 c_{j,k+1} +\textrm{h.c.} \Big] \  , \non\\    
	\mathcal{H}_2 &=\sum_{j,k}^{}\Big[c^\dagger_{j,k} \mathcal{T}_5 c_{j+1,k}+c^\dagger_{j,k} \mathcal{T}_6 c_{j,k+1} +\textrm{h.c.} \Big]\ , \quad  \mathcal{H}_3 =\sum_{j,k}^{}\Big[c^\dagger_{j,k} \mathcal{T}_7 c_{j+1,k}+c^\dagger_{j,k} \mathcal{T}_8 c_{j,k+1} +\textrm{h.c.} \Big]\ ,
\label{FMidHam}    
\end{align}
with
\begin{align}
	\mathcal{T}_1&= \frac{\mathcal{J}_0(A)}{2}\big(i t_1 \sigma_3\tau_1  + t_2 \sigma_0 \tau_3+\Delta \sigma_1 \tau_1\big)\ , \quad \mathcal{T}_2= \frac{\mathcal{J}_0(A)}{2}\big(i t_1 \sigma_0\tau_2  + t_2 \sigma_0 \tau_3-\Delta \sigma_1 \tau_1\big)\ , \non \\
    \mathcal{T}_3&= \frac{\mathcal{J}_1(A)}{2}\big( t_1 \sigma_3\tau_1  -i t_2 \sigma_0 \tau_3-i\Delta \sigma_1 \tau_1\big)\ , \quad \mathcal{T}_4= \frac{\mathcal{J}_1(A)}{2}\big( i t_1 \sigma_0\tau_2  + t_2 \sigma_0 \tau_3-\Delta \sigma_1 \tau_1\big)\ , \non \\
    \mathcal{T}_5&= \frac{\mathcal{J}_2(A)}{2}\big(-i t_1 \sigma_3\tau_1  - t_2 \sigma_0 \tau_3-\Delta \sigma_1 \tau_1\big)\ , \quad \mathcal{T}_6= \frac{\mathcal{J}_2(A)}{2}\big(i t_1 \sigma_0\tau_2  + t_2 \sigma_0 \tau_3-\Delta \sigma_1 \tau_1\big)\ , \non \\
    \mathcal{T}_7&= \frac{\mathcal{J}_3(A)}{2}\big( -t_1 \sigma_3\tau_1  +i t_2 \sigma_0 \tau_3+i\Delta \sigma_1 \tau_1\big)\ , \quad \mathcal{T}_8= \frac{\mathcal{J}_3(A)}{2}\big( i t_1 \sigma_0\tau_2  + t_2 \sigma_0 \tau_3-\Delta \sigma_1 \tau_1\big)\ .
\end{align}
\end{widetext}
For this intermediate frequency regime, we depict the numerical results obtained from the diagonaliztion of $\tilde{H}_F$ [see Eq. (\ref{midHam})] in Fig.~\ref{2Dlaser2}. The quasi-energy spectrum is 
shown around $0$ (modulo $\omega$) in Fig.~\ref{2Dlaser2} (a). Here, by considering the modulo operation, we have transmuted the modes appearing at quasi-energy $n\omega$ ($n \in \mathbb{Z}$) 
to quasi-energy $0$. The four $0$ (modulo $\omega$) dressed modes are depicted by red dots. One can compute the LDOS for the $0$ (modulo $\omega$) dressed modes to identify their corner localization. It is evident from Fig.~\ref{2Dlaser2} (b) that the dressed corner modes are located at the four corners of the system. Afterward, one can calculate the quadrupole moment for the 
$m^{\rm th}$ Floquet zone as $Q_{xy,m}$ to topologically characterize these dressed corner modes. We observe that $Q_{xy,m}=0.5$ (mod 1) for all the Floquet zone \ie $m=0,1,2,3$.

\subsection{Perturbation kick in three dimensions}
So far, we have focussed our discussion on 2D systems to generate FHOTI employing different kinds of time-dependent periodic modulations. Now, we switch to a 3D system, where we can generate 
FSOTI hosting 1D gapless hinge modes and Floquet TOTI (FTOTI) manifesting 0D corner modes. We employ the following static 3D FOTI Hamiltonian~\cite{Zhang2009,Slager2013,Nag2021} to dynamically generate a 3D FHOTI,
\begin{eqnarray}~\label{Eq:2TI_Hamiltonian}
	H^{\rm stat}_{\rm FOTI} &=& t \sum_{j=1}^3\Gamma_j \; S_j + \Gamma_4 \; [(m-6 t_0)+ 2 t_0 \sum^3_{j=1}C_j] \nonumber \\
	&\equiv& \sum^{4}_{j=1} N_j (\vect{k}) \; \Gamma_j,
\end{eqnarray}
where we define $S_j\equiv\sin(k_ja)$, $C_j\equiv\cos(k_ja)$; $\vect{k}$ being the crystal momentum. We set the lattice constant $a=1$ for convenience. The $4 \times 4$ mutually anticommuting $\vect{\Gamma}$ matrices are given as $\Gamma_i=\sigma_1  \tau_i$ with $i=1,2,3$, and $\Gamma_4= \sigma_3  \tau_0$. The Pauli matrices $\tau_\mu$ and $\sigma_\mu$ act on the orbital and spin degrees of freedom, respectively. The band inversion takes place at the high-symmetry $\Gamma=(0,0,0)$-point of the Brillouin zone when $0<\frac{m}{t_0}<4$. One obtains gapless 2D surface states with a crossing around zero-energy, ensured by anti-unitary particle-hole symmetry $P= \sigma_2 \tau_2 {\mathcal K}$ and unitary spectral symmetry $C=\Gamma_5 (= \sigma_2  \tau_0)$, with ${\mathcal K}$ being the complex conjugation operator. The LDOS associated with the zero-energy states of the Hamiltonian $H^{\rm stat}_{\rm FOTI}$ [Eq.~(\ref{Eq:2TI_Hamiltonian})] obeying OBC in all three directions is depicted in Fig.~\ref{3DPertrbationKick1} (a). One can identify that the gapless states are populated at the surfaces of the system.

\subsubsection{\rm FSOTI}
To dynamically generate a 3D FSOTI, we introduce a periodic kick in FOTI by the following WD mass term as~\cite{Nag2021}
\begin{equation} \label{Eq:periodickick3}
	V(\tilde{t}) =  V_1 \; \Gamma_{5} \sum^{\infty}_{r=1} \; \delta \left( \tilde{t}- r \; T \right),
\end{equation}
where $V_{1}= \sqrt{3} \Delta_1 \left( \cos k_1-\cos k_2 \right)$ is the four-fold $C_4$ rotation symmetry breaking WD mass and $T$ is the period of the kick and $\tilde{t}$ represent time and $r \in \mathbf{Z}$. One obtains the static counterpart of this driven system by considering a Hamiltonian $H^{\rm stat}_{\rm SOTI}=H^{\rm stat}_{\rm FOTI} + V_1 \Gamma_5$. Here, $V_1 \Gamma_5$ opens up a gap in the surface states of the FOTI. However, $V_1$ changes its sign across the $xz$ and $yz$ surfaces and one obtains 1D gapless hinge modes at the intersecting region along the $z$-axis~\cite{schindler2018}. Now we showcase the consequences when a static FOTI [Eq.~\eqref{Eq:2TI_Hamiltonian}] is periodically kicked by such a WD mass [Eq.~\eqref{Eq:periodickick3}]. On this account, we construct the Floquet operator associated with the driven Hamiltonian $H^{\rm stat}_{\rm FOTI}+V(\tilde{t})$ as
\begin{align}~\label{Eq:Floquetoperator}
	U(\vect{k},T) &= {\rm TO}\left( \exp \left[ -i\int_0^T \left[H^{\rm stat}_{\rm FOTI} + V(\tilde{t}) \right] d\tilde{t} \right] \right)  \ , \nonumber \\
	&= \exp(-i H^{\rm stat}_{\rm FOTI} \; T) \; \exp(-i   V_{1} \; \Gamma_5 ) \ .
\end{align}
One can obtain a closed-form effective Floquet Hamiltonian $H_{\rm Flq}=i \ln (U(\vect{k},T))/T$ in the high-frequency limit by employing the limits: $T \to 0$, $V_1 \to 0$, but $V_1/T$ is finite. 
Thus one obtains
\begin{equation}~\label{Eq:FloquetHamiltonian-TI}
	H^{\rm HF}_{\rm Flq} = \sum^4_{j=1} N_j ({\bf k}) \Gamma_j + V_{1} \; \sum^4_{j=1} N_{j} ({\bf k}) \Gamma_{j5} + \frac{V_1}{T} \; \Gamma_{5}\ .
\end{equation}
The effective Floquet Hamiltonian ($H_{\rm Flq}$ or $H^{\rm HF}_{\rm Flq}$) respects the antiunitary particle-hole symmetry $P$.

\begin{figure}[]
	\centering
	\includegraphics[width=0.47\textwidth]{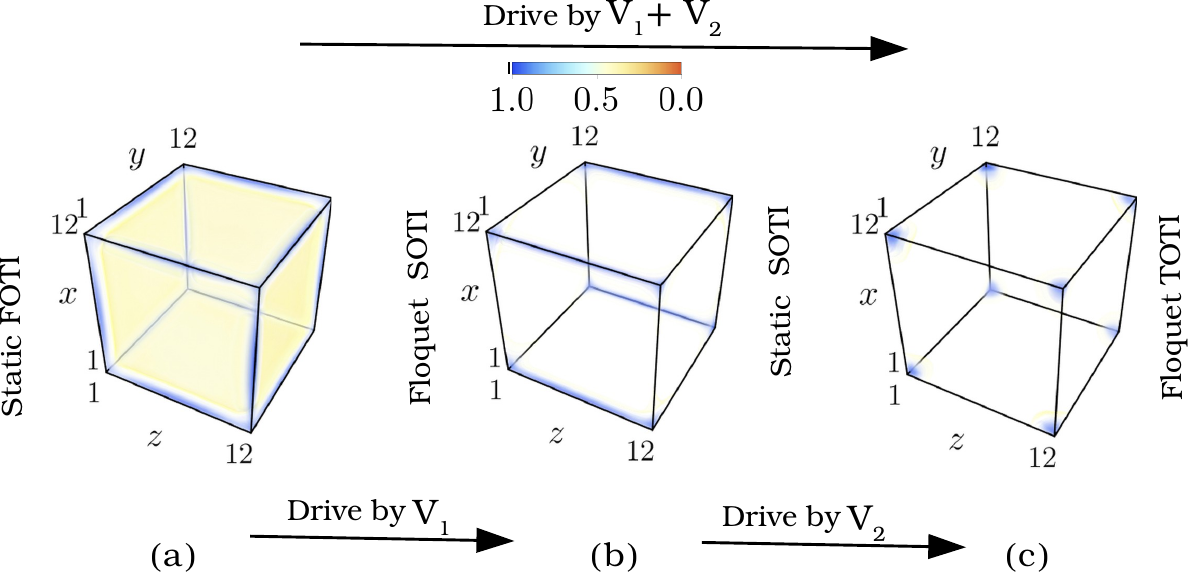}
	\caption{LDOS corresponding to (a) surface states of a static FOTI (involving eight-dimensional $\Gamma$ matrices), (b) dynamic hinge modes of Floquet SOTI (by periodically driving a static FOTI 
	by $V_1$), and (c) dynamic corner modes of Floquet TOTI (by periodically driving a static FOTI by $V_1$ and $V_2$) is demonstrated. We choose the model parameters as $t=t_0=1$, $m=2$, 
	$\Delta_1=\Delta_2=0.3$, and $\omega=10\gg t, t_0$ (assuring the high-frequency regime), see Eq.~(\ref{Eq:periodickick4}). We also find dynamic corner modes as in panel (c), when a \emph{static} 
	SOTI is periodically driven by only $V_2$. Throughout LDOS is normalized by its maximum value. This figure is adapted from Ref.~\cite{Nag2021} \copyright APS.
	}
	\label{3DPertrbationKick1}
\end{figure}

To demonstrate the footprints of the 1D gapless hinge modes of the FSOTI, we tie up with numerical analysis. We diagonalize the Floquet operator in Eq.~(\ref{Eq:Floquetoperator}), satisfying: $U(\vect{k},T) \; |\phi_n \rangle =\exp(i \mu_n T) \; |\phi_n \rangle$; with $|\phi_n \rangle$ being the Floquet quasi-states having quasi-energy $\mu_n$. We employ OBC in all three directions and depict the LDOS associated with the zero-quasienergy states in Fig.~\ref{3DPertrbationKick1} (b). One can identify that the zero-quasienergy states are sharply populated along the $z$-directed hinges such that 
the modes are localized along $x$- and $y$-direction but propagating along the $z$-direction. Thus, one obtains an FSOTI hosting gapless dispersive 1D hinge modes by periodically kicking static FOTI with a WD mass term. Similarly, by breaking the $C_4$ symmetry about the $x$ or $y$ axis, one can procure the hinge modes along the same axis.

\subsubsection{\rm FTOTI}
Having demonstrated the generation of 3D FSOTI, we now investigate the generation of 3D FTOTI hosting 0D corner modes using the periodic kick. Note that, one can find a maximum of five mutually anticommuting $4 \times 4$ hermitian matrices. However, we have already exhausted all five matrices to generate a 3D SOTI (static or dynamic). Thus, to proceed with the generation of the next hierarchical phase of HOTI \ie TOTI, by gapping out the $z$ hinges, we need to introduce $8 \times 8$ mutually anticommuting $\vect{\Gamma}$ matrices~\cite{Nag2021}. One can find seven mutually anticommuting $8 \times 8$ hermitian matrices. We employ the following representation of seven mutually anticommuting $8 \times 8$ Hermitian $\vect{\Gamma}$ matrices as
\begin{align}~\label{Eq:gammatriceseightdimensional}
	\Gamma_1&=\Sigma_1 \sigma_1  \tau_1, \Gamma_2=\Sigma_1 \sigma_1  \tau_2, \Gamma_3= \Sigma_1\sigma_1  \tau_3, \Gamma_4= \Sigma_1 \sigma_3  \tau_0, \nonumber \\
	\Gamma_5&=\Sigma_1 \sigma_2  \tau_0, \;\; \Gamma_6=\Sigma_3 \sigma_0 \tau_0, \;\; \Gamma_7=\Sigma_2 \sigma_0 \tau_0.
\end{align}
We denote Pauli matrices $\Sigma_\mu$ operate on the sublattice degrees of freedom. Afterward, we periodically kick a static FOTI [Eq.~(\ref{Eq:2TI_Hamiltonian})] by two WD masses~\cite{Nag2021}
\begin{align}~\label{Eq:periodickick4}
	V(\tilde{t}) =  \bigg( V_1  \; \Gamma_{5} + V_2 \; \Gamma_6 \bigg) \: \sum^{\infty}_{r=1} \; \delta \left( \tilde{t} - r \; T \right),
\end{align}
where $V_2= \Delta_2 (2 \cos k_3 - \cos k_1 -\cos k_2)$.

Before exploring the driven case, let us first try to understand the static counterpart which is given by the following Hamiltonian
\begin{equation}~\label{eq:3dcornerHamil}
	H^{\rm stat}_{\rm TOTI}= H^{\rm stat}_{\rm FOTI} + V_1  \; \Gamma_{5} + V_2 \; \Gamma_6 \equiv H^{\rm stat}_{\rm SOTI} + V_2 \; \Gamma_6.
\end{equation}
One can observe that when $V_2=0$, the above Hamiltonian represents a SOTI hosting 1D gapless hinge modes.  While the second WD mass term $V_2$ vanishes only along eight body-diagonal $(\pm 1, \pm 1, \pm 1)$ when $V_1$ is present. Thus, the addition of the second mass term $V_2$ gaps out the hinge states of SOTI and supports eight zero-energy localized corner modes. Hence, the system becomes a TOTI. The antiunitary particle-hole symmetry $P=\Sigma_1 \sigma_2 \tau_2 {\mathcal K}$ and unitary chiral symmetry $C=\Gamma_7$ are responsible for the stabilization of the corner states 
at zero energy. 

Now, we demonstrate a periodic kick in two WD mass terms [Eq.~\eqref{Eq:periodickick4}], to dynamically generate the FTOTI hosting corner modes at zero quasi-energies. With this periodic kick, 
one can construct the Floquet operator as $U(\vect{k},T)=\exp(-i H^{\rm stat}_{\rm FOTI} \; T) \exp(-i V_1 \Gamma_5 -i V_2 \Gamma_{6})$. The effective Floquet Hamiltonian is thus obtained from the Floquet operator, in the high-frequency limit (with $T, V_1, V_2 \to 0$ but $V_1/T$ and $V_2/T$ are finite) as 
\begin{align}~\label{Eq:FloquetHamiltonian2}
	H^{\rm HF}_{\rm Flq} = \sum^4_{j=1} N_j (\vect{k}) \left[  \Gamma_j +  V_1 \Gamma_{j5} +  V_2 \Gamma_{j6} \right]
	+ \sum_{j=1}^2\frac{V_{j}}{T} \; \Gamma_{j+4}\ .
\end{align}
The spectral symmetry of $H^{\rm HF}_{\rm Flq}$ is respected by the antiunitary operator $P=\Sigma_1 \sigma_2 \tau_2 {\mathcal K}$. To obtain the footprints of a 3D FTOTI, we diagonalize the Floquet operator by employing OBC along all three directions and compute the LDOS at quasi-energy zero. The corner-localization of the zero-quasienergy modes is evident from Fig.~\ref{3DPertrbationKick1} (c).

On the other hand, one may also start from a static SOTI model Hamiltonian $H^{\rm stat}_{\rm SOTI}$ and periodically kick the system employing the WD mass $V_2$. In this case, the Floquet operator reads as $U(\vect{k},T)= \exp(-i H^{\rm stat}_{\rm SOTI} \; T) \exp(-i V_2 \Gamma_{6})$. The corresponding effective Floquet Hamiltonian in the high-frequency limit reads as
\begin{equation}~\label{Eq:FloquetHamiltonian_TI3}
	H^{\rm HF}_{\rm Flq} = \sum^5_{j=1} N_j (\vect{k}) \Gamma_j + V_{2} \; \sum^5_{j=1} N_{j} (\vect{k}) \Gamma_{j6} + \frac{V_2}{T} \; \Gamma_{6}\ ,
\end{equation}
where $N_5(\vect{k}) \equiv V_1$. This driven setup also provides us with an FTOTI hosing localized corner modes [see Fig.~\ref{3DPertrbationKick1}]. To topologically characterize this FTOTI, we compute the octupolar moment $O_{xyz}$ using the Floquet operator (see Sec.~\ref{Sec:3DTOTIstatic} and Refs.~\cite{wheeler2018many,kang2018many} for details). The system exhibits $O_{xyz}=0.5$ (mod 1), which topologically characterizes a TOTI.

\subsection{Step drive in three dimensions} \label{Sec:Stepdrive3D}

In the previous subsection, we discuss periodic kick protocol to generate 3D Floquet HOTI hosting only $0$-quasienergy modes. Here, we showcase another driving protocol- step-drive protocol to generate both $0$- and $\pi$-modes in 3D systems. This driving protocol resembles that we have implemented in Sec.~\ref{Sec:StepDrive2D} to generate 2D FSOTI. Nevertheless, we introduce the 
two-step drive as follows~\cite{ghosh2021systematic}
\begin{align}
	H_{3 \rm D}&= J_1' h_{1, 3 \rm D}(\vect{k}) \  ,   \quad \quad t \in \Bigg[0,\frac{T}{2}\Bigg] \ , \nonumber \\
	&=J_2' h_{2, 3 \rm D}(\vect{k}) \  , \quad \quad t \in \Bigg(\frac{T}{2},T \Bigg] \ .
	\label{ch5_drive1}
\end{align}
As before, we define two dimensionless parameters $(J_1,J_2)=(J_1' T, J_2' T)$ to control the drive protocol. At the first driving step, we use $h_{1, 3 \rm D}(\vect{k})=\mu _x \sigma_z $ whereas in the send step, we employ $h_{2, 3 \rm D}(\vect{k})=\left( \cos k_x + \cos k_y + \cos k_z \right) \mu _x \sigma_z+ \sin k_x \mu_x \sigma_x s_x  + \sin k_y\mu_x \sigma_x s_y + \sin k_z \mu_x \sigma_x s_z + \alpha \left( \cos k_x - \cos k_y \right)\mu_x \sigma_y + \beta ( 2 \cos k_z -\cos k_x - \cos k_y ) \mu_z$ \cite{Nag2021,ghosh2021systematic}. We tune the dimensionless parameters $\alpha$ and $\beta$ to systematically generate the hierarchy of FSOTI and FTOTI phases. The terms associated with $\alpha$ and $\beta$ denote the WD mass terms as discussed in the previous subsection. The three Pauli matrices $\vect{\mu}$, $\vect{\sigma}$, and $\vect{s}$ act on sublattice $(A,B)$, orbital $(a,b)$, and spin $(\uparrow,\downarrow)$ degrees of freedom, respectively.

Before proceeding with the dynamical case, we first discuss the static analog of the model. In particular, we consider the following Hamiltonian 
\begin{align}
	H_{3 \rm D}^{\rm Static}(\vect{k})=& J_1' h_{1, 3 \rm D}(\vect{k}) + J_2' h_{2, 3 \rm D}(\vect{k}) \ . \label{ch5_3DStatic}
\end{align}
One can tune $\alpha$ and $\beta$ to observe the hierarchy of first-, second-, and third-order topological phases. By setting $\alpha=\beta=0$, we turn off both the WD mass terms. In this case, the Hamiltonian $H_{3 \rm D}^{\rm Static}(\vect{k})$ exhibits a FOTI hosting gapless 2D surface states in the strong TI phase when $0< \lvert J_1' \rvert < 3 \lvert J_2' \rvert $,~\cite{Zhang2009,Nag2021,Ghosh2021PRB}. As we set $\alpha$ to a non-zero value, the surface states of the 3D FOTI are gapped out, but we observe gapless states across the hinges of the system, designating a SOTI phase~\cite{benalcazar2017,schindler2018,Nag2021,Ghosh2021PRB}. On the other hand, for both $\alpha,\beta \neq 0$ the system becomes a TOTI manifesting localized 0D corner modes~\cite{benalcazar2017,Nag2021,Ghosh2021PRB}.

Moving our attention to the dynamical case, one can follow a similar procedure as discussed in Sec.~\ref{Sec:StepDrive2D} and obtain the phase boundaries in terms of $J_1$ and $J_2$ as
\begin{equation}\label{ch5_phasestep3D}
	\frac{3 \lvert J_2 \rvert}{2}   = \frac{\lvert J_1 \rvert}{2} + n \pi \ ,
\end{equation}
where $n \in \mathbb{Z}$. The phase diagrams, in the $J_1 \mhyphen J_2$ plane, remain unaltered for FSOTI and FTOTI in 3D~[see Figs.~\ref{3DStep1} (a) and \ref{3DStep2} (a), respectively] as Eq.~(\ref{ch5_phasestep3D}) is independent of $\alpha$ and $\beta$. Thus, similar to the 2D case [see Sec.~\ref{Sec:StepDrive2D}], the phase diagram is divided into four parts in the 
$J_1 \mhyphen J_2$ plane- R1, R2, R3, and R4.

\subsubsection{\rm FSOTI}

\begin{figure}[]
	\centering
	\includegraphics[width=0.49\textwidth]{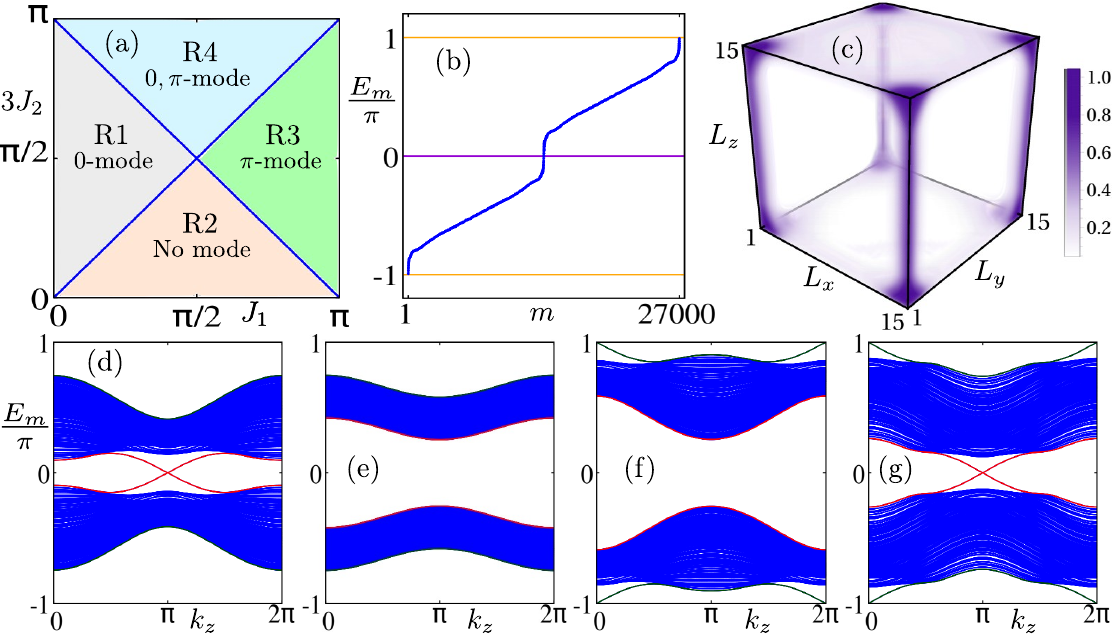}
	\caption{(a) We demonstrate the phase diagram in the $J_1 \mhyphen J_2$ plane for a 3D FSOTI. (b) Quasi-energy spectrum, $E_m$ for the finite size system, is shown as a function of the 
	state index, $m$ for R4. (c) The corresponding LDOS is depicted considering quasi-states with $E_m=0,~\pm \pi$ in R4. The quasi-energy spectra $E_m$, considering rod geometry along $k_z$, 
	corresponding to R1, R2, R3, and R4, are shown in panels (d), (e), (f), and (g), respectively. The chosen model parameters are: $(J_1, 3 J_2)=\left[ \left(\frac{\pi}{4},\frac{\pi}{2}\right),~\left(\frac{\pi}{2},\frac{\pi}{4}\right),~\left(\frac{3\pi}{4},\frac{\pi}{2}\right),~\left(\frac{\pi}{2},\frac{3 \pi}{4}\right) \right]$ for R1, R2, R3, and R4, respectively. This figure is adapted from Ref.~\cite{ghosh2021systematic} \copyright APS.
	}
	\label{3DStep1}
\end{figure}

We set $\alpha \neq 0$ and $\beta=0$, to realize the FSOTI phase. The corresponding phase diagram is illustrated in Fig.~\ref{3DStep1} (a). We diagonalize the Floquet operator employing OBC in all three directions. We depict the quasi-energy spectra in Fig.~\ref{3DStep1} (b), when the system resides in the R4 phase with both the Floquet modes present. One can clearly identify the existence of both $0$ and $\pi$-modes. Afterward, we demonstrate the LDOS corresponding to quasistates with $E_m=0,~\pm \pi$ for R4, in Fig.~\ref{3DStep1} (c). We observe that both the $0$- and $\pi$-modes are populated along the $z$-hinges of the system. However, to notice the dispersive nature of the hinge modes, we resort to rod geometry \ie PBC in one direction ($z$-direction) and OBC in the remaining two directions ($x$ and $y$-direction). We show the rod geometry quasi-energy spectra in Figs.~\ref{3DStep1} (d), (e), (f), and (g) for  R1, R2, R3, and R4, respectively. Thus, employing this two-step periodic drive, one can generate 3D FSOTI exhibiting gapless modes around quasi-energy $0$ and $\pi$.

\subsubsection{\rm FTOTI}
\begin{figure}[]
	\centering
	\includegraphics[width=0.49\textwidth]{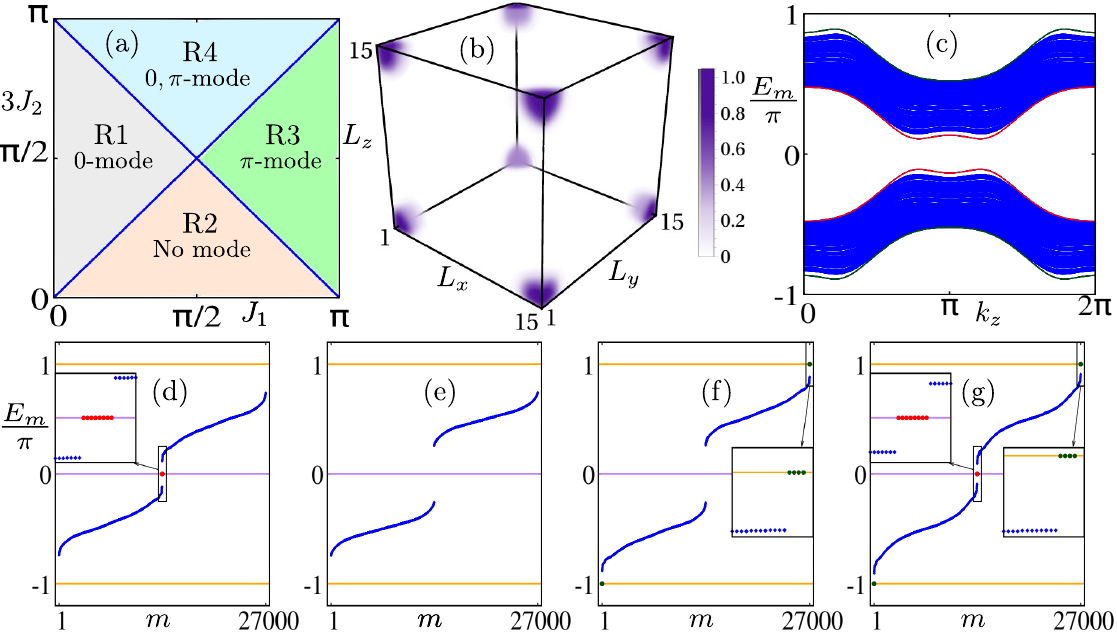}
	\caption{(a) We illustrate the phase diagram in the $J_1 \mhyphen J_2$ plane for the 3D FTOTI. (b) Corner localized modes corresponding to quasi-energies $E_m=0,~\pm \pi$ are shown via LDOS for the finite-size system in R4. (c) Quasi-energy spectrum $E_m$ for this system, considering rod geometry, is depicted as a function of $k_z$. This manifests that the hinge modes (both $0$ and $\pm \pi$) are gapped out. The quasi-energy spectra $E_m$, for a finite size system and as a function of the state index $m$, corresponding to R1, R2, R3, and R4 are shown in panels (d), (e), (f), and (g), respectively. We choose the same model parameter values as mentioned in Fig.~\ref{3DStep1}. This figure is adapted from Ref.~\cite{ghosh2021systematic} \copyright APS.
	}
	\label{3DStep2}
\end{figure}

When we set both $\alpha$ and $\beta$ to a non-zero value, we can obtain the FTOTI phase. We depict the phase diagram to highlight the FTOTI in Fig.~\ref{3DStep2} (a) choosing the $J_1 \mhyphen J_2$ plane. We show the LDOS, computed at $E_m=\pm \pi$ for R4, as a function of the system dimensions in Fig.~\ref{3DStep2} (b). From the latter, one can clearly identify that both the $0$- and $\pi$-modes are localized at the eight corners of the system. Both the bulk and hinge states of an FTOTI exhibit a gapped structure. We demonstrate the quasi-energy spectrum employing rod geometry \ie we consider OBC along $x$ and $y$-directions, PBC along $z$-direction in Fig.~\ref{3DStep2} (c) (for R4). It is evident that the hinge modes are gapped around both $0$ and $\pm \pi$ quasi-energy. 
Here, the second WD mass term $\beta (2 \cos k_z-\cos k_x -\cos k_y)$ is responsible for gapping out the hinge states~\cite{Nag2021}. Then, we consider OBC along all three directions and depict the corresponding quasi-energy spectra in Figs.~\ref{3DStep2} (d), (e), (f), and (g) for R1, R2, R3, and R4, respectively. The $0$ and $\pi$ corner states are denoted by red and green dots respectively 
in the insets.

\subsection{Mass kick in three dimensions}

\begin{figure}[]
	\centering
	\includegraphics[width=0.35\textwidth]{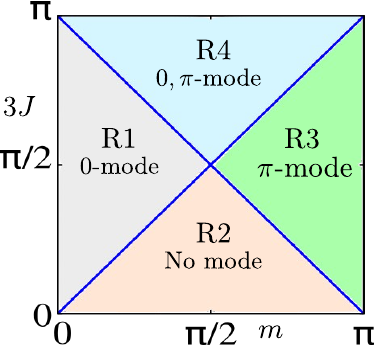}
	\caption{(a) Phase diagram for the kick-driving protocol is depicted in the $m \mhyphen J$ plane for a 3D FSOTI/FTOTI. This figure is adapted from Ref.~\cite{ghosh2021systematic} \copyright APS.
	}
	\label{3Dmasskick}
\end{figure}

For the mass kick drive protocol, we consider the Hamiltonian $\mathcal{H}_{3 \rm D}$ between two successive kicks. Afterward, we introduce the driving protocol as~\cite{ghosh2021systematic}
\begin{equation} 
	m_0(t)=m \ h_{1, 3 \rm D} \sum_{r=1}^{\infty} \delta(t-rT) \ , 
	\label{ch5_kick1sys3D}
\end{equation}
where, $m$, $t$, and $T$ signify the kicking parameter's strength, time, and time-period of the drive, respectively. Following the periodic kick protocol, one can construct the Floquet operator, $U_{3 \rm D}(\vect{k},T)$  as
\begin{align}
	U_{3 \rm D}(\vect{k},T)&=\overline{\rm TO} \exp \left[-i\int_{0}^{T}dt\left(\mathcal{H}_{3 \rm D}({\vect{k}})+m_0(t) \right)\right] \ ,  \nonumber \\
	&= \exp(-i \mathcal{H}_{3 \rm D}({\vect{k}}) T)~\exp(-i m~h_{1, 3 \rm D})\ .
	\label{ch5_fomasskick3D}
\end{align}
We choose $\mathcal{H}_{3 \rm D}({\vect{k}})=J' h_{2, 3 \rm D}(\vect{k})$. Employing this mass kick protocol, one can obtain the gap-closing condition as
\begin{align}
	3 \lvert J \rvert =& \lvert m \rvert  + n \pi \ . \label{phasekick3D}
\end{align}
The nature of the phase boundary remains invariant for all the topological orders \ie FSOTI and FTOTI. The corresponding phase boundaries are shown in Fig.~\ref{3Dmasskick}.

For the sake of completeness, we discuss the numerical results obtained for the mass kick drive protocol in three dimensions. When $\alpha \neq 0$ but $\beta=0$, the system exhibits FSOTI hosting gapless 1D hinge modes. The corresponding dispersive signature is manifested in the quasi-energy spectrum employing rod geometry [similar to Figs.~\ref{3DStep1}~(d), (e), (f), and (g)]. On the other hand, the FTOTI phase is obtained, when $\alpha,\beta \neq 0$. In the FTOTI phase, the trace of the corner localized modes (at $E_{m}=0, \pm\pi$) are found in the quasi-energy spectrum for a system obeying OBC in all three directions [akin to Figs.~\ref{3DStep2} (d), (e), (f), and (g)]. However, we do not intend to show these behavior in order to avoid repetition.

\section{Discussions and Outlook}  \label{sec5}
In this section, we discuss the current challenges and future opportunities. To be precise, the detrimental effects of disorder, interaction, temperature  are very much relevant for the generation and further stabilization of a dynamic topological phase. There exist plenty of relevant tunable parameters for the driven systems that cannot only minimize the above challenges but also open up  future opportunities for technological solutions. In this regard, throughout this review, we have treated our systems as clean \ie disorder-free and at zero temperature. Although, this might not be the case in a practical situation. Nevertheless, corner modes should be persistent in the presence of weak disorder and at finite temperature, with the disorder and temperature scale being smaller than the bulk bandgap of the system. However, the effect of strong disorder, with its strength being comparable to the bandwidth and in the presence of external irradiation, can be fascinating as far as the Anderson insulator phase is concerned. Also, the generation of corner modes at quasi-energy $\omega/2$ employing periodic laser irradiation still remains an open question. Moreover, temperature/ heat up effect and temporal disorder effects are unexplored so far in the field of Floquet topological insulator to the best of our knowledge. 

The interplay of disorder and Floquet engineering can result in the intriguing phase called Floquet topological Anderson insulator~\cite{TitumPRB2017,titumPRX2016,torresPRB2019} where a periodically driven trivial system becomes topological in the presence of disorder with anomalous Floquet boundary modes. On the other hand, the static Anderson phase has been contextualized for the higher-order topological system in the current literature~\cite{LiPRL2020,arakiPRB2019,ybyangPRB2021,jhwangPRL2021,yshuPRB2021}. However, realizing the \textit{Floquet higher-order topological Anderson insulator} (FHOTAI) phase with anomalous boundary modes is still lacking in the literature. The main challenge here is to find an appropriate topological invariant in real space that can serve the purpose of the marker indicating the topological phase transition in the dynamical sector.  Thus, there is a need for developing a topological invariant that can topologically characterize an anomalous higher-order Floquet topological mode in the presence of strong disorder (FHOTAI phase). 

In another direction, it has already been established that the topological phases can be realized in a system with substantial electro-electron correlation (Hubbard interaction). Especially, extensive investigation has been performed in the case of first-order topological systems, in particular the Kane-Mele-Hubbard model~\cite{RachelPRB2010,ShunLiPRL2011,Hohenadler2013}.  The presence 
of strong Coulomb interaction gives rise to a new phase called ``topological Mott insulator''. Therefore, in the context of higher-order topological systems, the effect of strong correlations can be an interesting research direction considering systems like breathing Kagome lattice~\cite{Ezawakagome}, modified Haldane model on a hexagonal lattice~\cite{BaokaiPRBL2021}, and multiorbital models 
like BHZ model with a $\mathcal{C}_4$-symmetry breaking WD mass term~\cite{schindler2018}, etc. The expected primary outcome can be the generation of higher-order topological Mott insulators 
with anti-ferromagnetic/ other magnetic order, topological phase transition induced via strong correlation, etc. Although, understanding and characterization of any interaction-driven topological phase 
via appropriate topological invariant remains a challenging task.

Moreover, another intriguing aspect of the time-dependent system, apart from generating the dynamical topological phases, is the robustness of the topological states. In this direction, 
the fate of the topologically protected states of FOTIs has been investigated upon applying time-dependent perturbation~\cite{FedorovaLight2019}. It has been observed that although the topological characteristics of the bulk remain intact, the edges tend to depopulate~\cite{FedorovaLight2019}. Thus, it would be interesting to perform a similar study based on the higher-order systems and investigate whether the higher-order modes also exhibit similar behavior compared to the first-order modes.

\section{Experimental developments and material perspectives}  \label{sec6}
The quest for HOTIs has surged for the last five years after their theoretical discovery. The 2D quadrupolar and 3D octupolar topological phases have been realized experimentally employing metamaterial platforms mainly phononic crystals~\cite{serra2018observation}, acoustic systems~\cite{xue2019acoustic,ni2019observation,Experiment3DHOTI.aSonicCrystals,Ni2020}, electric-circuit setups~\cite{imhof2018topolectrical}, and photonic lattice~\cite{PhotonicChen,PhotonicXie,mittal2019photonic} etc. Noticeably, 3D SOTI hosting gapless dispersive 1D hinge modes has been proposed in 
SnTe (based on first-principle calculations)~\cite{schindler2018} and realized experimentally in bismuth hallide~\cite{schindler2018higher,Aggarwal2021}, Bi$_{0.92}$Sb$_{0.08}$~\cite{Aggarwal2021}, bismuth-bromide (Bi$_4$Br$_4$)~\cite{Experiment3DHOTI.VanDerWaals,ShumiyaHOTI2022}, and WTe$_2$~\cite{LeeNatureComm2023} etc. However, there is not much experimental evidence of 
2D SOTI and 3D TOTI based on solid state systems. Thus the experimental development in the field of HOTI is still in its infancy as far as the real material platform is concerned.

On the other hand, the theoretical analysis of Floquet engineering of band topology has also accelerated a few tantalizing experimental observations~ \cite{WangScience2013,Mahmood2016,McIver2020,Jotzu2014,Wintersperger2020,Peng2016,fleury2016floquet,RechtsmanExperiment2013,Maczewsky2017,Bao2022}. However, in a solid-state or real material platform, the observation of Floquet states are limited to time-resolved and angle-resolved photoemission spectroscopy~(TrARPES) study of Floquet-Bloch states in Bi$_2$Se$_3$ surface~\cite{WangScience2013,Mahmood2016} and detection of light-induced quantum anomalous Hall effect in graphene~\cite{McIver2020}. Moreover, in a few cold-atomic systems, such Floquet states have been experimentally predicted~\cite{Jotzu2014,Wintersperger2020}. As far as the experimental generation of FHOTIs is concerned, there has been one proposal based on accoustic setup utilizing the 
step drive protocol to realize this phase~\cite{experimentFloquetHOTI}. Therefore, note that the setups that we discuss in the context of FHOTIs in this present review are yet to be realized from the experimental point of view in real materials and thus open up a plethora of future experimental research directions. However, the development of sophisticated experimental techniques to investigate 
and engineer time-dependent systems would facilitate the generation of FHOTIs in a real material platform.

\section{Summary and Conclusions}  \label{sec7}
To summarize, in this topical review, we have provided a introduction to the new emerging field of HOTI and their driven counterpart FHOTI in quantum condensed matter physics. In these intriguing $n^{{\rm{th}}}$-order HOTI phases, gapless/localized boundary modes reside on $(d - n)$-dimensional boundaries of a $d$-dimensional system, rather gapless $(d - 1)$-dimensional boundary modes in FOTIs. We showcase various periodic driving protocols to generate FHOTI in two and three dimensions. We demonstrate that some of the driving protocols also allow us to realize both $0$ as well as the dynamical $\pi$ higher-order modes. The latter do not have any static analog. In particular, we discuss the development in which a perturbation kick protocol, a two-step drive protocol, a mass-kick protocol, and laser irradiation are individually employed to generate 2D FSOTI hosting 0D corner modes. In three dimensions, we introduce similar kind of periodic drive protocols \ie perturbation kick, step drive, and mass kick protocol to achieve the higher-order Floquet phases. Interestingly, one can generate two topological orders in three dimensions- FSOTI hosting 1D gapless dispersive hinge modes and FTOTI manifesting 0D localized corner modes. On the experimental side, fabricating different setups of fermionic systems to realize HOTI as well as their driven counterpart FHOTI still remains a challenging task. From the application point of view, the topological propagating 1D hinge modes can be potential candidate towards future spintronics applications and 0D localized corner modes can be the building block for the fault-tolerant topological quantum computation. All in all, there are still surprises in store as we probe deeper into the realm of static/driven topological quantum matter along with substantial experimental challenges.

\subsection*{Acknowledgments}
A.K.G. and A.S. acknowledge SAMKHYA: High-Performance Computing Facility provided by Institute of Physics, Bhubaneswar, for numerical computations. A.K.G. and A.S. acknowledge Ganesh C Paul and T.N. acknowledges Bitan Roy, Vladimir Juricic, Debmalya Chakrabarty, Saptarshi Mandal, Sudarshan Saha, and Rodrigo Arouca for stimulating discussions on higher-order topological systems. T.N. would also like to thank Andras Szabo, and Dumitru Calugaru for their technical support. T.N. is deeply grateful to his Ph.D. supervisor Prof. Amit Dutta, whose sudden demise is a great loss to all of us.

\bibliographystyle{apsrev4-2mod}
\bibliography{bibfile.bib}

\end{document}